\pdfoutput=1

\documentclass{sig-alternate} 
\usepackage{mathptmx} 

\usepackage{graphicx}
\newcommand{\ignore}[1]{}
\usepackage[pass]{geometry}
\usepackage{fancyhdr}
\usepackage[normalem]{ulem}
\usepackage[hyphens]{url}

\usepackage[sort,nocompress]{cite}

\usepackage[final]{microtype}
\usepackage{flushend}
\usepackage[bookmarks=false]{hyperref}

\usepackage{alltt}
\usepackage{epsfig}
\usepackage{graphicx}
\usepackage{enumitem}
\usepackage{amsmath}

\usepackage{amssymb}
\usepackage[normalem]{ulem}
\usepackage{float}
\usepackage[]{algorithm2e}
\usepackage{lipsum}
\usepackage{balance}
\usepackage{tikz}
\usepackage{calc}
\usepackage{xcolor, colortbl}
\usepackage{multirow}
\usepackage{pifont}
\usepackage{adjustbox}
\usepackage{array}

\usepackage{tcolorbox}
\usepackage{amssymb,amsmath}

\usepackage{xfrac}
\usepackage{mdframed}

\usepackage{dblfloatfix}

\usepackage{color,soul}
\definecolor{light-gray}{gray}{0.91}
\definecolor{lightgreen}{RGB}{195, 233, 211}
\sethlcolor{lightgreen}

\soulregister\cite7
\soulregister\ref7
\soulregister\pageref7

\makeatletter
\def\SOUL@hlpreamble{%
	\setul{}{2.2ex}
	\let\SOUL@stcolor\SOUL@hlcolor
	\SOUL@stpreamble
}
\makeatother

\newcommand\thintilde{{\lower.92ex\hbox{\mathtt{\char`\~}}}}
\newcommand\thicktilde{{\lower.74ex\hbox{\texttt{\char`\~}}}}

\usepackage{amsfonts}
\usepackage{url}

\fancypagestyle{firstpage}{
	\fancyhf{}
	\setlength{\headheight}{50pt}
	
	\pagenumbering{arabic}
}

\usepackage{authblk}
\author[$\dag$]{Vinson Young}
\author[$\ddag$]{Zeshan Chishti}
\author[$\dag$]{Moinuddin K. Qureshi\vspace{-.1in}}
\affil[$\dag$]{Georgia Institute of Technology}
\affil[$\ddag$]{Intel Corporation}
\affil[ ]{\texttt {\{vyoung,moin\}@gatech.edu},zeshan.a.chishti@intel.com\vspace{-.15in}}
\title{
\vspace{-.3in}
\vspace{-.33in}
TicToc: Enabling Bandwidth-Efficient DRAM Caching \\ for both Hits and Misses in Hybrid Memory Systems
\vspace{-.13in}
\vspace{-.45in}
}

\begin{document}
\maketitle
\thispagestyle{firstpage}
\pagestyle{plain}

\begin{abstract}

This paper investigates bandwidth-efficient DRAM caching for hybrid DRAM + 3D-XPoint memories. 3D-XPoint is becoming a viable alternative to DRAM as it enables high-capacity and non-volatile main memory systems. However, 3D-XPoint has several characteristics that limit it from outright replacing DRAM: 4-8x slower read, and even worse writes. As such, effective DRAM caching in front of 3D-XPoint is important to enable a high-capacity, low-latency, and high-write-bandwidth memory. There are currently two major approaches for DRAM cache design: (1) a Tag-Inside-Cacheline (TIC) organization that optimizes for hits, by storing tag next to each line such that one access gets both tag and data, and (2) a Tag-Outside-Cacheline (TOC) organization that optimizes for misses, by storing tags from multiple data lines together in a tag-line such that one access to a tag-line gets information on several data-lines. Ideally, we would like to have the low hit-latency of TIC designs, and the low miss-bandwidth of TOC designs. To this end, we propose a \textit{TicToc} organization that provisions both TIC and TOC to get the hit and miss benefits of both.

We find that naively combining both techniques actually performs worse than TIC individually, because one has to pay the bandwidth cost of maintaining both metadata. The main contribution of this work is developing architectural techniques to reduce bandwidth cost of accessing and maintaining both TIC and TOC metadata. We find that most of the update bandwidth is due to maintaining the TOC dirty information. We propose a \textit{DRAM Cache Dirtiness Bit} technique that carries DRAM cache dirty information to last-level caches, to help prune repeated dirty-bit updates for known dirty lines. We also propose a \textit{Preemptive Dirty Marking} technique that predicts which lines will be written and proactively marks the dirty bit at install time, to help avoid the initial dirty-bit update for dirty lines. To support PDM, we develop a novel PC-based \textit{Write-Predictor} to aid in marking only write-likely lines. Our evaluations on a 4GB DRAM cache in front of 3D-XPoint show that our TicToc organization enables 10\% speedup over the baseline TIC, nearing the 14\% speedup possible with an idealized DRAM cache design with 64MB of SRAM tags, while needing only 34KB SRAM.

\end{abstract}


\ignore{
This paper investigates channel-sharing DRAM + 3D-XPoint memories, along with their implications for DRAM cache management.
Industry is moving towards providing a DRAM cache and 3D-XPoint in the same channel for multiple reasons: to provide the illusion of a high-capacity low-latency memory, and to ensure each channel has DRAM behind it to maximize the bandwidth from each channel. However, while channel-sharing is effective for bus utilization, provisioning 3D-XPoint and DRAM in the same channel means that 
any DRAM cache maintenance operation such as miss-probes or installs will now come at a direct cost to bus bandwidth for memory. 
Prior miss-bandwidth for hit-latency tradeoffs that have been proven effective for stacked-DRAM caches no longer hold effective -- we need renewed analysis on DRAM caching techniques for channel-sharing DRAM caches.
}

\ignore{

This paper investigates channel-sharing DRAM + 3D-XPoint memories, along with their implications for DRAM cache management.
Industry is moving towards providing DRAM (as a cache) and 3D-XPoint in the same channel for multiple reasons: to provide the illusion of a high-capacity low-latency memory, and to ensure each channel has DRAM behind it to maximize the bandwidth possible from each channel. However, while channel-sharing is effective for bus utilization, provisioning 3D-XPoint and DRAM in the same channel means that 
any DRAM cache maintenance operation (install, miss probe) will now come at a direct cost to the bus bandwidth available for memory. Prior miss-bandwidth-for-hit-latency tradeoffs that have been proven effective for stacked DRAM caches may no longer hold effective for a channel-sharing setup. This motivates a renewed analysis on DRAM caching techniques for channel-sharing DRAM caches.

There are two major practical approaches for tags-in-DRAM DRAM caching:
(1) an Alloyed Metadata Organization (TIC) that optimizes for hits by storing tag next to each line, such that one access gets both tag and data, and (2) a Tag-Outside-Cacheline (TOC) that optimizes for misses by storing tags from multiple data lines together in a tag-line, such that one access to a tag-line gets information on several data-lines. 
We can provision both tags in a dual metadata approach to get the hit and miss benefits of both. However, we still find that a dual metadata approach spends significant bandwidth updating the aggregated-tags.
We propose a \textit{TicToc Organization (HMO)}  approach that is intelligently designed to get the hit and miss benefits of dual metadata, without paying significant costs of maintaining either.
We find the majority of update bandwidth is due to maintaining the dirty bit in TOC, as these bits are updated at L3 eviction time and have poor spatial locality.
We propose a \textit{DRAM Cache Dirtiness Bit} technique that carries DRAM cache dirty bit to last-level caches, to help prune repeated dirty-bit updates for known dirty lines. 
We propose a \textit{Preemptive Dirty Marking} technique that predicts which lines will be written and proactively marks the dirty bit early at install time, to help avoid even the first dirty-bit update for likely-to-be-written lines. We develop a novel signature-based \textit{Write-Predictor} to aid in classifying likely-to-be-written lines. Our evaluations on a 4GB DRAM cache in front of 3D-XPoint show that our HMO techniques enable 11\% speedup over a baseline TIC approach, nearing the 13\% speedup possible with an idealized DRAM cache design with 64MB of SRAM tags, all while needing only \textasciitilde 34KB of SRAM.

}

\section{Introduction}


As memory systems scale, non-volatile memories or NVMs (such as, 3D-XPoint~\cite{intel:3dxpoint}) are emerging as viable alternatives to DRAM. NVMs offer the advantages of higher bit density and the ability to retain data after power outages. 
However, NVMs also have significant limitations that prevent them from outright replacing DRAM in the memory hierarchy. For example, 3D-XPoint is reported to have 4-8x slower read, and even slower writes compared to DRAM~\cite{xpoint_latency2}. As such, future systems are likely to utilize hybrid memory systems~\cite{optane_dramcache,hybridpcm_1,hybridpcm_2,hybridpcm_3} consisting of both DRAM and 3D-XPoint.
We focus on the setup where DRAM is operated as a hardware-managed cache for 3D-XPoint based main memory, since such a setup enables applications to benefit from the lower latency and higher write-bandwidth of DRAM and the higher capacity of 3D-XPoint without relying on any software or OS support.

\begin{figure*}[htb] 
	\centering
    \vspace{-0.35 in}
    \includegraphics[height=1.75in]{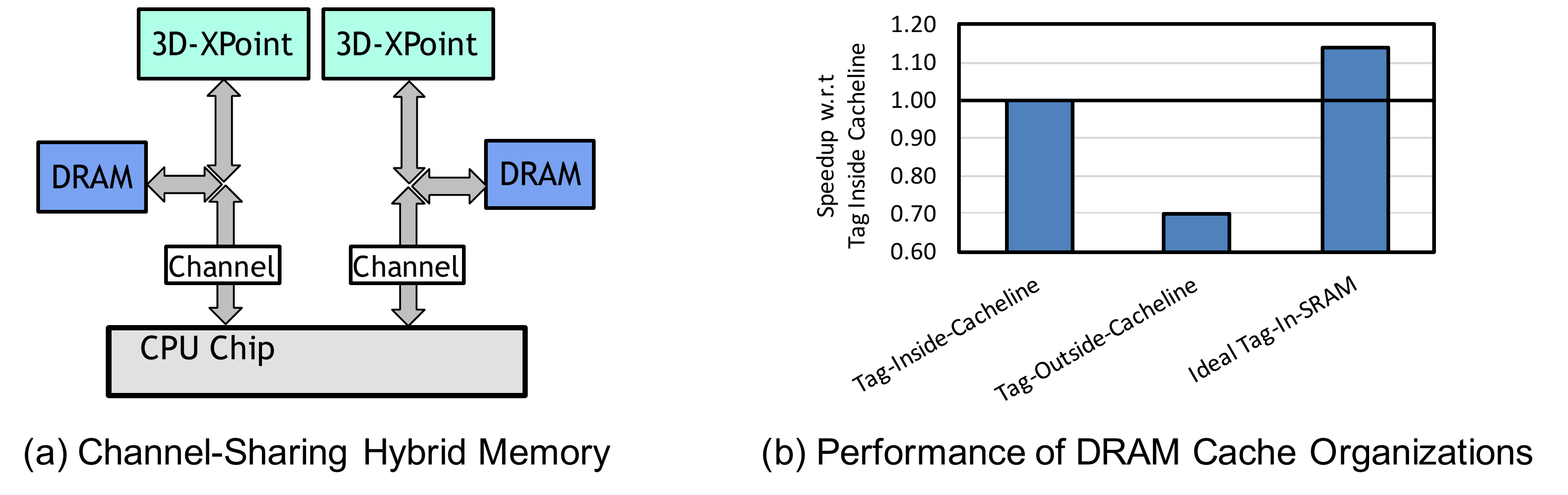}
    \vspace{-0.17 in}
    \caption{(a) Channel-Sharing Hybrid Memory, and (b) Performance of hit-optimized Tag-Inside-Cacheline (TIC)~\cite{Alloy}, miss-optimized Tag-Oustide-Cacheline (TOC)~\cite{timber}, and idealized Tag-In-SRAM, normalized to TIC.} 

    \vspace{-0.17 in}
	\label{fig:intro} 
\end{figure*}


Recently, there have been many works~\cite{LHCache,FootprintCache:ISCA2013:MICRO2014papers,Alloy,KNL,timber,simcache,unison} on architecting High Bandwidth Memory (HBM)~\cite{HBM} caches in front of traditional DRAM main memory~\cite{ddr4}.
These works target \textit{improving memory bandwidth} by migrating data between DRAM and HBM, and servicing most data at the higher internal/bus bandwidth of HBM. 
These works are effective due to HBM having dedicated higher-bandwidth channels/interfaces compared to commodity DDRx DRAM.
We would like to utilize the insights learned from these works to design effective DRAM caches in front of NVMs, such as 3D-XPoint. However, we note that there are significant differences in setup and goals for a DRAM+3D-XPoint hybrid memory as compared to a \ignore{traditional }HBM+DDRx hybrid memory.

First, in a 3D-XPoint based hybrid memory, 3D-XPoint and DRAM will be sharing the same channel interfaces~\cite{same_protocol}. Second, DRAM caches in front of 3D-XPoint target \textit{reducing read latency} and \textit{improving write bandwidth and endurance} of 3D-XPoint, by servicing most data at the lower latency and higher write bandwidth of DRAM. 
An added complexity is that the DRAM cache and 3D-XPoint are likely to sit behind the same channel~\cite{channel_sharing}, as depicted in Figure~\ref{fig:intro}(a). Such a set-up enables a balanced configuration where every channel has DRAM backing it. However, in such a channel-sharing set-up, bandwidth needed for maintaining DRAM cache state now comes directly at a cost to bus bandwidth available for memory. As such, there is a renewed need for bandwidth-efficient DRAM caches. We analyze prior DRAM caching approaches, highlight cases of bandwidth-inefficiency, and rigorously target the remaining bandwidth overheads to develop a bandwidth-efficientf DRAM cache suitable for DRAM + 3D-XPoint systems.




We start with a baseline hit-optimized \textit{Tag-Inside-Cacheline (TIC)} DRAM cache design~\cite{KNL,Alloy,BEAR}.
A TIC design organizes its DRAM cache as a direct-mapped cache with tags stored inside each cacheline, such that one access can retrieve both tag and data. \textit{TIC has good hit-latency}, as it can service cache hits in one DRAM access. However, TIC incurs bandwidth overhead on cache misses as it needs to probe the tag in DRAM in order to determine a miss. 
This approach of trading miss-bandwidth for hit-latency has been proven effective in situations where the cache has its own dedicated access channel, such as the HBM+DRAM hybrid memory in Intel's Knights Landing~\cite{KNL}.
However, in a channel-sharing setup, the miss probe bandwidth directly consumes available main memory bandwidth, resulting in bandwidth inefficiency. As we show in Figure~\ref{fig:intro}(b), there is a 14\% performance gap between TIC and an idealized Tag-In-SRAM approach.


An alternative approach to DRAM cache design is a miss-optimized \textit{Tag-Outside-Cacheline (TOC)} design~\cite{timber,simcache,unison}. A TOC design stores tags of multiple cachelines together in a tag-only-line, such that one access to a tag-line can obtain information for multiple cachelines at once. We can bring in these bundles of tags as needed, and cache them in a small tag cache (e.g., 32KB SRAM)~\cite{timber}. If the tag cache has high hit-rate, TOC can service most hits with one DRAM access, and misses to clean lines without a DRAM access. However, if the tag cache has low hit-rate, TOC may need two accesses to service a hit, and one access to service a miss.
As such, \textit{TOC  consumes lower bandwidth on misses} than TIC; however, it consumes higher bandwidth on hits due to separate tag and data read. 
Overall, as shown in Figure~\ref{fig:intro}(b), TOC approach performs worse than TIC due to bandwidth overheads.


We notice that TIC is good for hits, while TOC is good for misses -- one can perhaps combine both approaches to get both good hit and miss bandwidth. Fortunately, it is cheap to provision both metadata at once: TIC uses spare ECC bits~\cite{KNL}, and TOC needs to dedicate only 1.5\% of DRAM cache capacity to store metadata and a 32KB SRAM for a metadata cache~\cite{timber}. 
To decide when to use TOC or TIC, one can employ a hit/miss predictor~\cite{Alloy} that uses TIC for likely hits and TOC for likely misses. 
We call this proposal that provisions both TIC and TOC metadata as \textit{TicToc}.
Unfortunately, we find that naively combining TIC and TOC actually leads to worse performance than TIC by itself. This is because maintaining and updating TOC metadata bits consumes significant DRAM bandwidth. In order for \textit{TicToc} to be effective, we need ways to reduce TOC maintenance bandwidth.


TOC incurs bandwidth overheads for the following three cases: (i) tag-check on hits, ( ii) tag-update on installs, and (iii) dirty-bit-update on writebacks. Hit overhead is easily mitigated by additionally storing TIC metadata in \textit{TicToc}. Tag updates are generally inexpensive because they occur at miss time, and miss traffic usually has good spatial locality and therefore a high metadata-cache hit-rate. Dirty-bit updates, however, remain costly because they are carried out when dirty lines are evicted from an earlier level of cache. Such evictions have poor access locality and therefore low metadata-cache hit-rates. Hence, we identify dirty bit updates as the most significant bandwidth overhead for \textit{TicToc}.


To reduce dirty data tracking costs for TOC, we target the following two cases: initial write to a cache line, and repeated writes to the same cache line. For repeated writes, we propose to store a \textit{DRAM Cache Dirtiness} bit alongside the line in an earlier level of cache, to track the current dirty status of the line in the DRAM cache. On a writeback to DRAM cache, we only need to update the TOC if the line in the DRAM cache has changed from clean to dirty. However, many workloads write to lines only once. For such workloads, we propose \textit{Preemptive Dirty Marking} that predicts likely-to-be-written cache lines and proactively mark sthose lines as dirty in the TOC at install time. This avoids needing to update dirty information at eviction time, thereby avoiding metadata-cache misses. We develop a PC-based \textit{Write Predictor} that is 92\% accurate for our Preemptive Dirty Marking.

Even after solving for hit and miss bandwidth, when data has poor reuse, installing lines and updating TOC tag can become a major source of bandwidth overhead. To mitigate that problem, we develop a \textit{Write-Aware Bypassing} technique that reduces install and tag-update bandwidth, without increasing writes to write-constrained 3D-XPoint.


\vspace{.02in}

Overall our paper makes the following contributions:

\vspace{.02in}





{\setlength{\parindent}{0cm} {\bf Contribution-1:} This paper evaluates and rigorously targets the bandwidth overheads of prior DRAM-cache organizations. We find that we can combine two tag-storage methods with a \textit{TicToc} organization to obtain both good hit and good miss path. However, such an approach suffers significant bandwidth cost to maintain TOC dirty information on writes.} 

\vspace{.03in}


{\setlength{\parindent}{0cm} {\bf Contribution-2:} We develop two techniques to reduce the cost of tracking dirty information. \textit{DRAM Cache Dirtiness Bit} targets reducing cost of dirty-bit updates for repeated writes to the same location, via maintaining DRAM cache dirty information alongside the line in an earlier level of cache. And, \textit{Preemptive Dirty Marking} targets reducing cost of the initial dirty-bit update to a location, via predicting which lines are likely to be written to (with our \textit{Signature-based Write Predictor}) and preemptively setting the dirty-bit.}

\vspace{.03in}

{\setlength{\parindent}{0cm} {\bf Contribution-3:} To reduce install bandwidth while not increasing 3D-XPoint write traffic, we develop a Write-Aware Bypass technique. This technique bypasses most clean lines by default to save install bandwidth. And, it installs most dirty and predicted write-likely lines (to amortize metadata updates) to buffer writes to write-constrained 3D-XPoint.}

\vspace{.03in}

Overall, our proposed TicToc organization, enables 10\% speedup over TIC baseline, nearing the 14\% speedup of an idealized Tag-In-SRAM approach, while needing significantly less SRAM storage (34 KB vs. 64 MB).


\ignore{

Need capacity. 3d-XPoint way for capacity. but, high latency. HW managed DRAM caching. 

Recently, many works, but mainly organizations that uses DRAM cache primarily for BW (dedicated separate channel). This work targets DRAM caching for 3d-XP. 

Channels for DRAM, channels for NVM. Split channels with with dedicated DRAM, and dedicated NVM. but, if high hit-rate, can only utilize half total bus bandwidth. Fig1(a)

Instead, channel sharing has higher potential. Fig1(b)
Product has mentioned such an organization. No paper discussing benefits of conjoined hybrid memories. Our work focuses on analyzing the potential benefits and challenges of such an organization

Benefits: use of all channel bandwidth. Challenges: DRAM cache organization -- BW bloat directly consumes bus bandwidth of memory. Need careful design.

Our baseline DRAM cache approach, Alloy/KNL. Proven effective for hits, but costs for miss. need 

Alternative, tags together and cache such that one access gets information for multiple lines. Used in several papers. Tag cache hit, near ideal. Better potential for miss. Benefit depends on hit-rate of tag cache. but, worse for hits, because can potentially need two accesses to read. 

Can combine both for good hit and miss. But, actually worse off than Alloy, because saddled with cost of maintaining both tags. Need effective ways to reduce cost of maintaining both tags.

Deep-dive into cost of maintaining tags. TOC updated at miss+install has good spatial locality, small cost. Agg-dirty updated on eviction of dirty line from L3, which has poor spatial and temporal locality, high cost. Need effective ways to reduce Agg-dirty update cost.

DRAM Cache Dirtiness Bit, for repeated writes 
Preemptive Dirty Marking + Write Predictor, for initial writes.
In total, Dual-tag with dirty-update optimizations enable 11\% speedup, within 2\% of ideal SRAM tag approach with much less storage.

Contrib:
1. Conjoined Hybrid Memories. No prior work on this topic. Analyze benefits (DRAM behind each channel for high memory throughput). And difficulties (DRAM cache BW overheads more impactful.

2. Optimize hit/miss with dual-tags. but cost of maintaining both tags. 

3. Find major cost of maintaining dual-tags is updating dirty-bit. DCD bit for repeated writes. Write-predictor (initializes dirty-bit as dirty at install time to avoid separate dirty-bit update). 92\% accurate. 

}

\begin{figure*}[htb]
		\centering
   \vspace{-0.35in}
	\centerline{\includegraphics[height=2.15in]{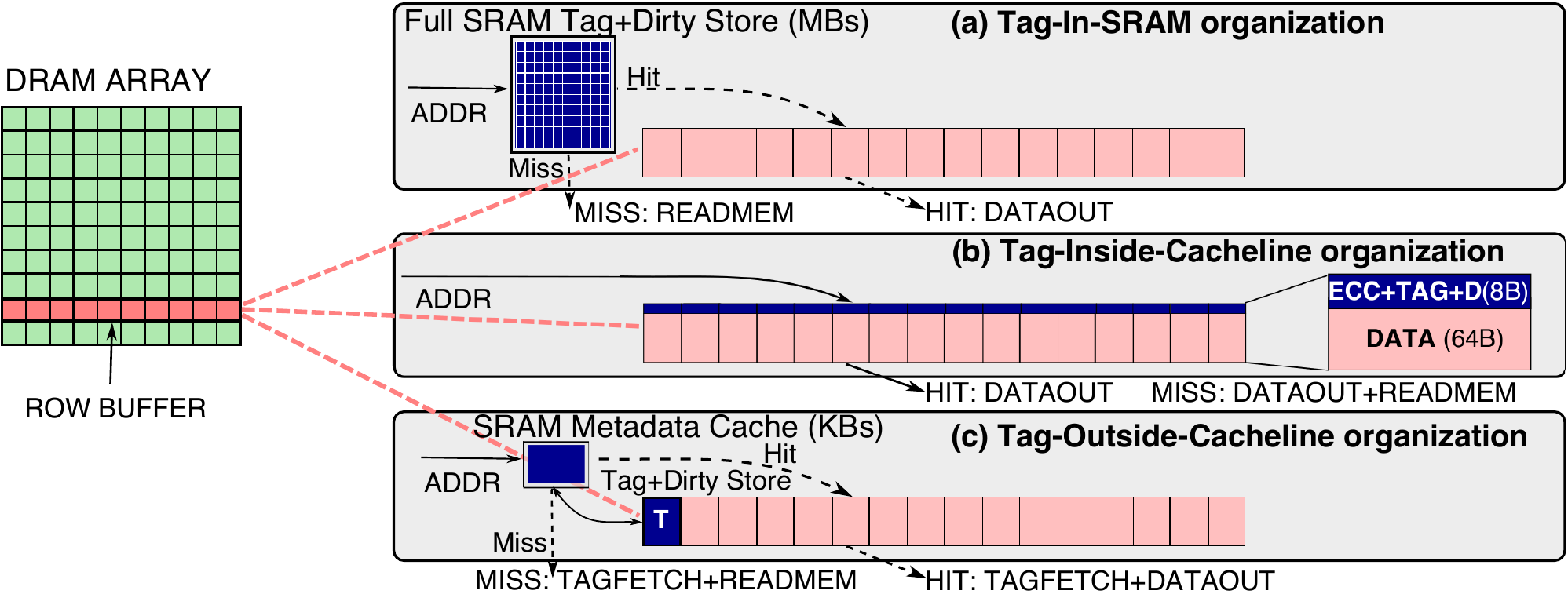}}
	\vspace{-0.15in}
	\caption{DRAM cache organization and flow for (a) idealized Tag-In-SRAM, (b) hit-latency-optimized Tag-Inside-Cacheline (TIC)~\cite{Alloy}, and (c) miss-bandwidth-optimized Tag-Outside-Cacheline (TOC)~\cite{timber}.}
    \vspace{-0.15 in}
	\label{fig:caches}
\end{figure*}

\newpage

\section{Background and Motivation}





DRAM caches are important for enabling heterogeneous memory systems to have the effective latency and bandwidth of one memory technology, and the capacity of another; however, there are several challenges in designing DRAM caches. 
A DRAM cache design has to balance multiple goals. 
First, it should minimize the SRAM storage needed for DRAM cache maintenance.
Second, it should minimize cache hit latency. Third, it should minimize miss latency. Fourth, it should provide high hit-rate. Lastly, it should try to minimize total bandwidth costs for maintaining DRAM cache state.

It is desirable to organize DRAM caches at the granularity of a cache line to efficiently utilize cache capacity, and to minimize the consumption of main memory bandwidth~\cite{FootprintCache:ISCA2013:MICRO2014papers}.
A key challenge in designing a large line-granularity cache is deciding where to store the tag and dirty-bit metadata.
For a moderately-sized 4GB DRAM cache with 64B lines, there would be 64 million lines. Even if each metadata required 8 bits (6 tag, 1 dirty, 1 valid bit), this would result in 64MB storage for metadata.
Next, we discuss the various options for DRAM cache metadata management, and their implications for SRAM storage cost and bandwidth consumption.

\begin{table}[htb]
\vspace{-.15in}
\caption{Bandwidth of DRAM Cache Organizations --\newline $\rho$ denotes Metadata-Cache Miss Probability}
\label{table_bw}
\centering
\setlength{\tabcolsep}{2.3pt}{
\begin{tabular}{|c|c|c|c|}
\hline
Organization  & SRAM & TIC & TOC \\ 
(SRAM Cost)       & (\textgreater 20MB) & (\textless 1KB) &   (\textasciitilde 32KB) \\ \hline \hline
Hit                 & 1               & 1                       & 1 + {\color{red}\textbf{$\rho$}}    \\ \hline
Miss + Evict-Clean  & 0               & {\color{red}\textbf{1}} & 0 + {\color{red}\textbf{$\rho$}}\\ \hline
Miss + Evict-Dirty  & 1               & 1                       & 1 + {\color{red}\textbf{$\rho$}}    \\ \hline
\end{tabular}
}
\vspace{-.1in}
\end{table}

\subsection{Tag In SRAM}

A costly method to design high performance DRAM caches is to simply maintain all of the tag and dirty bits in on-chip SRAM, and query the on-chip SRAM metadata to determine hit or miss, in a \textit{Tag-In-SRAM} approach, shown in Figure~\ref{fig:caches}(a). Assuming 1-byte metadata per cache line, such an approach would require 64MB of SRAM storage for a 4GB cache (\textgreater 20MB with sectoring~\cite{FootprintCache:ISCA2013:MICRO2014papers,sector}). Table~\ref{table_bw} shows the DRAM bandwidth consumption for such an approach. SRAM metadata is queried first to determine hit or miss. A hit can be serviced with one DRAM access to data. A miss can be serviced without a DRAM access to data. However, in preparation for installing the newly accessed line, the cache would need to perform an eviction of the resident line. If the resident line were clean, the location could be directly overwritten. However, if the resident line were dirty, the resident line would need to be read before writeback to memory. Hence, miss with eviction of a clean line costs 0 bandwidth, and miss with eviction of a dirty line costs 1 bandwidth. Such a design represents the minimum DRAM bandwidth needed for DRAM cache maintenance, and an upper-bound for performance. We aim to achieve Tag-In-SRAM performance at low SRAM cost.

\subsection{Tag Inside Cacheline}

To reduce SRAM storage costs, one could store tags inside each line in DRAM~\cite{KNL,Alloy,BEAR} in a \textit{Tag-Inside-Cacheline (TIC)} approach, shown in Figure~\ref{fig:caches}(b). 
TIC optimizes for hit-latency by using a direct-mapped design and storing tag inside each data-line such that one access can retrieve both tag and data.
Direct-mapped organization enables the controller to know which location to access, without waiting for tags. 

Table~\ref{table_bw} shows the bandwidth of such an approach. 
Hits are serviced with one DRAM access that retrieves both tag and data: in case of a tag match, the attached data can be used to service the request. However, misses also need to access tag in DRAM.
As such, TIC is effective for hit-latency, but consumes extra bandwidth on misses. This approach of trading miss-bandwidth for hit-latency has been proven effective in commercial products~\cite{KNL}, and, as such, we use the TIC organization~\cite{Alloy} as our baseline. 


\textbf{Setup:} We store metadata alongside data in unused ECC bits similar to Intel's Knights Landing ~\cite{KNL}. TIC additionally employs a small hit-miss predictor to guide when to access cache+memory either in a parallel or serial manner (needs \textless 1KB SRAM storage overhead). 
We additionally include bandwidth-reducing enhancements from Chou et al.~\cite{BEAR}, such as DCP to reduce writeback probe.

\subsection{Tag Outside Cacheline}
\label{ssec:our_TOC}

Another option with reduced SRAM storage costs, is to store metadata lines in a separate area of DRAM and bring them in as needed in a \textit{Tag-Outside-Cacheline (TOC)}\cite{timber,simcache,unison} approach, shown in Figure~\ref{fig:caches}(c). To determine hit or miss, TOC first accesses a metadata line to get tag+dirty information for the requested data line, then routes the request appropriately to DRAM cache or to memory. Of note, each of these metadata lines actually stores tag+dirty information of several adjacent data lines. An enhanced design~\cite{timber} proposes to cache the metadata lines in a small metadata cache to avoid repeated accesses to the same metadata line, and would amortize metadata lookup if there is 
spatial locality. 
Table~\ref{table_bw} shows the bandwidth consumption of such an approach. In case of a metadata-cache hit, TOC performs similar to idealized Tag-In-SRAM. For a metadata-cache miss, TOC spends additional bandwidth to access the metadata. Overall, TOC has the potential for reducing miss bandwidth, but it can suffer from significant bandwidth overhead when the metadata-cache has poor hit rate (due to poor spatial locality). 


\textbf{Setup:} We assume 1-byte metadata (6 tag, 1 dirty, 1 valid bits), and 64 tags stored in each metadata entry. The metadata are stored in a separate part of DRAM, consisting of 64MB out of the 4GB DRAM capacity. Recently accessed metadata are stored in a 512-entry metadata cache, which requires 32KB of SRAM. Note that the metadata cache is sized to capture only spatial locality and not the working set of the DRAM cache, which would need megabytes of SRAM. 

\textbf{Optimizing for Latency:} In the case of metadata-miss in the metadata-cache, we want to avoid the latency for serialized tag + cache-data access, as well as the latency for serialized cache-data + memory access. We employ a direct-mapped organization and a hit-miss predictor~\cite{Alloy} for latency and bandwidth considerations. If predicted hit, we access tag + cache-data in parallel to save latency (direct-mapped organization dictates only one possible location for data), and serially access memory only if prediction is wrong to save memory bandwidth. If predicted miss, we access tag + memory in parallel for latency, and serially access cache-data only if prediction is wrong to save DRAM cache bandwidth.

\subsection{Insight: Combine Metadata Approaches}


A TIC approach has good \textit{hit-latency}, but suffers from extra miss bandwidth. Whereas, a TOC approach has good \textit{miss bandwidth} but incurs extra hit bandwidth. Our key insight is that if one could use TIC for hits and TOC for misses, then one could potentially achieve both good hit and miss bandwidth. 

We note that provisioning metadata for both TIC and TOC is relatively inexpensive: TIC simply uses spare ECC bits~\cite{KNL}, and TOC needs to dedicate only \textasciitilde 1.5\% of DRAM cache capacity to store metadata lines and employs 32KB SRAM for its metadata cache~\cite{timber}. However, we need an effective design that can use TIC for hits and TOC for misses.

We notice that we can use hit/miss predictor~\cite{Alloy} to help guide when to use TIC or TOC. For predicted hits, we can directly access the line with TIC. For predicted misses, we can consult the metadata-line / metadata-cache in TOC to help avoid miss probes. We call this proposal \textit{TicToc}. Unfortunately, we find that naively combining both approaches actually leads to worse performance than TIC individually. This is because maintaining TOC tag and dirty bits consumes substantial bandwidth. To complete our design, we need to develop effective solutions to reduce maintenance bandwidth for TOC. We discuss methodology before proposed design.






\ignore{

\subsection{DRAM Cache Performance \& Bandwidth}

A TIC approach is optimized under high hit-rate circumstances, whereas an TOC approach is optimized under high spatial locality assumptions. We show the bandwidth consumption of these approaches to understand where these approaches suffer.

\begin{figure}[htb] 
	\centering
	\vspace{-0.10 in}
    \includegraphics[height = 1.3in]{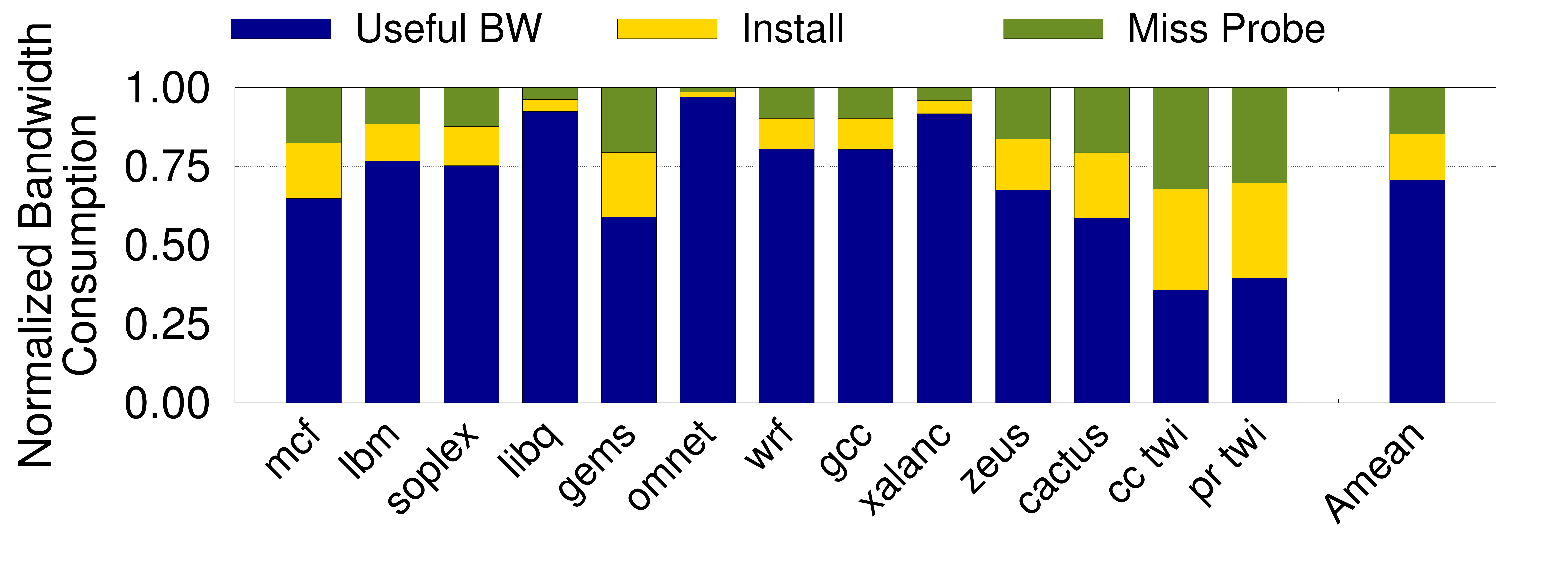}
    \vspace{-0.40 in}
	\caption{Breakdown of bus bandwidth consumption of the hit-optimized Distributed-Tags~\cite{Alloy} approach. Workloads with low hit-rate waste significant bandwidth.}
     \vspace{-0.1 in}
	\label{fig:tim3} 
\end{figure}

Figure~\ref{fig:tim3} and Figure~\ref{fig:tim4} show the proportion of bus bandwidth used for useful operations, install operations, and assorted maintenance operations, for Distributed-Tag~\cite{Alloy} and Aggregated-Tag~\cite{timber} DRAM cache organizations, respectively. Useful operations include 3D-XPoint Read and Write, and DRAM Cache Hit and Writeback. Install operations refer to cache installs, which are important for improving hit-rate but cost bandwidth to write the line to DRAM. Lastly, Maintenance operations refer to bandwidth-wasting operations used to confirm a line is not resident: miss probes in the case of distributed-tag, and aggregated-tag access / update in the aggregated-tag approaches. Overall, we find that prior tags-in-DRAM approaches often waste significant bandwidth to maintain the cache (e.g., \textit{pr twi} spends most of the bus bandwidth maintaining cache state and has poor performance).

\begin{figure}[htb] 
	\centering
	\vspace{-0.10 in}
    \includegraphics[height = 1.3in]{GRAPHS/bw_T.pdf}
    \vspace{-0.40 in}
	\caption{Breakdown of bus bandwidth consumption of the miss-optimized Aggregated-Tags~\cite{timber} approach. Workloads without spatial locality incur tag-update cost.}
     \vspace{-0.1 in}
	\label{fig:tim4} 
\end{figure}

\textbf{Our approach:} We employ a Dual-Tag approach to obtain both good hit and miss-paths, and additionally solve for Aggregated-Tag maintenance costs to enable a low-cost high-performance DRAM cache controller.

}


\ignore{
Motivation
    Common BW to any DRAM cache.
        Useful BW
            Hit. Writes (for write buffering). 
        Necessary for operation
            Install is necessary for now.
            If install over dirty resident line -- need to read before write to memory.
    Alloy -- overheads
        Miss probe.
            If install over clean resident line -- could have directly written without read.
        BEAR
            miss probe reduction (neighbor tag): not applicable
            writeback probe (DCP bit): used in baseline
            bypass-90\%: discussed in later section. can be improved.
    TIMBER -- overheads
        Tim-access 
            Hit, miss+install,
        Tim-update
            Install, writes for dirty bit.
            
    Alloy+TIMBER
        Tim-access
            miss+install (spatial locality)
        Tim-update
            install (spatial locality), writes (no spatial locality. solve).
            
    e.g. GRAPH workloads. Miss probes, install BW, few hits. No DRAM actually works much better. Dynamic turning-off possible, but still too expensive -- also, fails for mixed workloads. Rather, work on reducing all bandwidth overheads for always-on approach.
}

\ignore{
\subsection{TOC Metadata Implementation}
\label{ssec:our_TOC}
There are substantial latency and bandwidth implications in how the metadata-cache miss path of TOC is handled.
We find the original TOC documentation~\cite{timber} lacking in details of how the serialization of metadata, data, and memory accesses could be avoided. We discuss our latency-optimized and bandwidth-trimmed implementation. 

On L3 miss, we first query the metadata cache to determine L4 hit or miss. In the case of metadata-hit in metadata-cache, TOC performs with ideal latency and bandwidth of 1 access for hit, 0 for miss+clean-evict, and 1 for miss+dirty-evict. In the case of metadata-miss in metadata-cache, we want to avoid the latency for serialized tag + cache-data access, as well as the latency for serialized cache-data + memory access. We employ a direct-mapped organization and a hit-miss predictor~\cite{Alloy} for latency and bandwidth considerations. If predicted hit, we access metadata + cache-data in parallel for latency (direct-mapped organization dictates only one possible location for data), and serially access memory only if prediction is wrong to save memory bandwidth. If predicted miss, we access metadata + memory in parallel for latency, and serially access cache-data only if prediction is wrong to save DRAM cache bandwidth.
}

\section{Methodology}
\subsection{Framework and Configuration}
\label{subsection:conf}

We use USIMM~\cite{USIMM}, an x86 simulator with detailed memory
system model.  We extend USIMM to include a DRAM cache.
Table~\ref{table:config} shows the configuration used in our study.
We assume a four-level cache hierarchy (L1, L2, L3 being on-chip SRAM
caches and L4 being off-chip DRAM cache). All caches use 64B line
size. We model a virtual memory system to perform virtual to physical
address translations. The baseline L4 is a 4GB DRAM-cache\cite{KNL}, which is direct-mapped and places tags with data in unused ECC bits. 
The parameters of our DRAM cache are based on DDR4 DRAM technology~\cite{ddr4}.
The main memory is based on 3D-XPoint\cite{xpoint_latency2,optane_latency,intel:3dxpoint}:
the read latency is \textasciitilde 6X, the write latency is \textasciitilde24X that of DRAM, and there are 64 rowbuffers each 256B in size.

\begin{table}[ht]
\vspace{-.18in}
  \begin{center}
      \caption{System Configuration}
          \vspace{0.05 in}
\renewcommand{\arraystretch}{.80}
\setlength{\extrarowheight}{2pt}{
      \begin{tabular}{|l|l|}  \hline 

Processors & 8 cores; 3.0GHz, 4-wide OoO \\
Last-Level Cache & 8MB, 16-way \\ \hline

\multicolumn{2}{|c|}{\bf DRAM Cache } \\ \hline

Capacity               &    4GB         \\
Bus Frequency          &     1000MHz (DDR 2GHz) \\
Configuration       &  1 channel, 64-bit bus, shared \\
Aggregate Bandwidth              &      16 GB/s, shared with Memory \\ 
tCAS-tRCD-tRP-tRAS     &      13-13-13-30 ns \\ \hline

\multicolumn{2}{|c|}{\bf Main Memory (3D XPoint) } \\ \hline

Capacity               &    64GB        \\
Bus Frequency          &     1000MHz (DDR 2GHz) \\
Configuration       &  1 channel, 64-bit bus, shared \\
Aggregate Bandwidth              &      16 GB/s, shared with DRAM \\ 
tCAS-tRCD-tRP    &      4-80-0 ns \\
tRAS-tWR     &          96-320 ns \\ \hline

      \end{tabular}
}
      \label{table:config}
\vspace{-0.27 in}

      \end{center}
\end{table}


\subsection{Workloads}
\label{subsection:workloads}

We use a representative slice of 2-billion instructions selected by
PinPoints~\cite{pinpoint}, from benchmark suites that include SPEC
2006~\cite{SPEC2006} and GAP~\cite{GAP}.  
For SPEC, we pick a subset of high memory intensity workloads that have at least 2 L3 misses per thousand instructions (MPKI).
The evaluations execute benchmarks in rate mode, where all
eight cores execute the same benchmark. In addition to rate-mode
workloads, we also evaluate 21 mixed workloads, which are created
by randomly choosing 8 of the 17 SPEC workloads. Table~\ref{table:benchmarks}
shows L3 miss rates, and memory footprints for the 8-core rate-mode workloads in our study. 

We perform timing simulation until each benchmark in a workload executes at least 2 billion instructions.  We use weighted speedup to measure aggregate performance of the workload normalized to the baseline and report geometric mean for the average speedup across all the 17 workloads (11 SPEC, 2 GAP, 4 MIX). We provide key performance results for additional 17 SPEC-mixed workloads in Section~\ref{ssec:mixed}.


\begin {table}[ht]

\vspace{-.18 in}
\caption{Workload Characteristics}
\begin{center}
\vspace{0.05 in}
\renewcommand{\arraystretch}{.70}
\setlength{\extrarowheight}{2.4pt}{
        \begin{tabular}{|c|c|c|c|} \hline

Suite & Workload & L3 MPKI & Footprint \\ \hline \hline

\multirow{10}{*}{SPEC} 
& mcf & 101.14 & 13.4 GB \\ \cline{2-4}
& lbm & 49.3 & 3.2 GB  \\ \cline{2-4}
& soplex & 35.3 & 1.8 GB \\ \cline{2-4}
& libq & 30.1 & 256 MB \\ \cline{2-4}
& gems & 29.1 & 6.4 GB  \\ \cline{2-4}
& omnet & 29.0 & 1.2 GB  \\ \cline{2-4}
& wrf & 10.4 & 1.1 GB \\ \cline{2-4}
& gcc & 7.6 & 1.5 GB \\ \cline{2-4}
& xalanc & 7.4 & 1.5 GB \\ \cline{2-4}
& zeus & 7.0 & 1.6 GB \\ \cline{2-4}
& cactus & 6.5 & 2.6 GB \\ \hline
\ignore{
& lbm & 49.3 & 3.2 GB  \\ \cline{2-4}
& soplex & 35.3 & 1.8 GB \\ \cline{2-4}
& libq & 30.1 & 256 MB \\ \cline{2-4}
& omnet & 29.1 & 1.2 GB  \\ \cline{2-4}
& leslie & 22.1 & 623 MB \\ \cline{2-4}
& tonto & 11.8 & 50 MB \\ \cline{2-4}
& wrf & 10.4 & 1.1 GB \\ \cline{2-4}
& gcc & 7.6 & 1.5 GB \\ \cline{2-4}
& zeus & 7.0 & 1.6 GB \\ \cline{2-4}
& cactus & 6.5 & 2.6 GB \\ \cline{2-4}
& dealII & 1.7 & 481 MB \\ \hline
}
\multirow{2}{*}{GAP} & cc twitter & 116.8 & 9.3 GB  \\ \cline{2-4}
& pr twitter & 126.6 & 15.3 GB    \\ \hline

        \end{tabular}
}
\vspace{-0.2 in}
\vspace{-0.1 in}
      \label{table:benchmarks}
    \end{center}
\end{table}


\vspace{-.1in}
\begin{figure*}[htb] 
	\centering
    \vspace{-0.3 in}

    \includegraphics[width=6.5in]{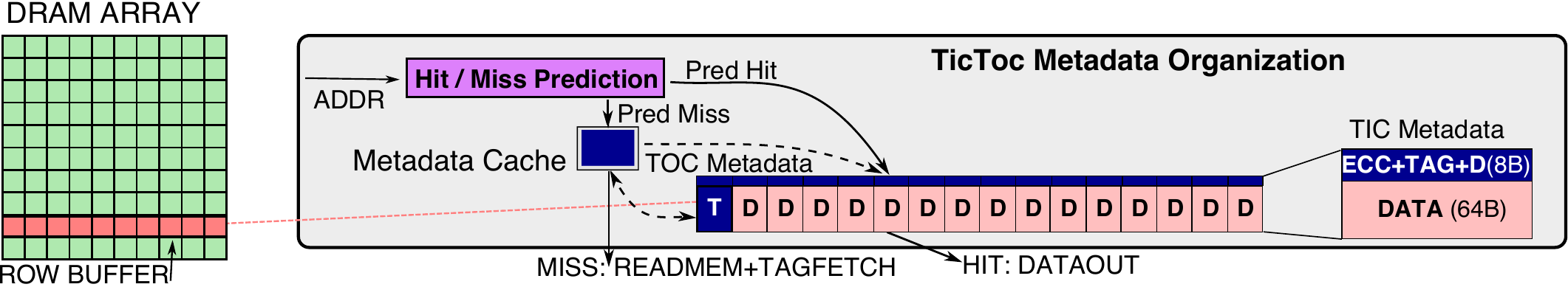}
    \vspace{-0.1 in}
    \caption{TicToc Metadata Organization queries hit/miss predictor to use TIC metadata for hits and TOC metadata for misses. TicToc enables good hit latency, and good hit/miss bandwidth.}
    \vspace{-0.12 in}
	\label{fig:tim_hybridtag} 
\end{figure*}

\newpage

\section{TicToc Design}


DRAM caches need metadata to confirm if a line is cache resident or not (tag bits), and if the resident line is the most up-to-date copy (dirty bit). 
Tag-Inside-Cacheline (TIC) organizations are optimized for hits as one access gets both metadata and data, but can suffer for misses as misses still need to access DRAM for metadata.
In contrast, Tag-Outside-Cacheline (TOC) organizations are optimized for misses as one metadata access gets residency and dirty information for multiple lines; however, such approaches suffer from needing to frequently query and update TOC metadata. Ideally, we want the hit-path of TIC, the miss-path of TOC, all without paying significant cost to access and maintain TOC metadata. This section is organized as follows: we describe how to provision and effectively use both TIC and TOC metadata in a TicToc organization, describe how to reduce TOC metadata maintenance costs,
and show effectiveness of our design.


\subsection{TicToc Metadata Organization}

Figure~\ref{fig:tim_hybridtag} shows metadata organization of our \textit{TicToc} design. TicToc provisions TIC metadata -- tag-bits and dirty-bit are stored inside the cacheline in unused ECC bits, similar to commercial implementation~\cite{KNL}. In addition, TicToc provisions TOC metadata -- metadata is stored in dedicated metadata lines corresponding to 1.5\% of DRAM capacity, and cached as needed in a
32KB on-chip metadata cache. While provisioning both TIC and TOC metadata is relatively cheap, the complexity lies in utilizing TIC and TOC metadata appropriately to save on bandwidth for both hits and misses.

\subsubsection{TicToc Operation} 
Figure~\ref{fig:tim_hybridtag} shows the operation of TicToc. Ideally, we want to use TIC metadata for hits and TOC metadata for misses. Our key insight is that one can use hit/miss prediction~\cite{Alloy,Sim:MICRO2012} to help guide when to use which metadata. Hit/miss predictors have been primarily used to hide the serialization latency that can occur from waiting on last-level cache response before sending memory access. They work by predicting which cache accesses are likely to miss, and sending both cache and memory requests in parallel to avoid serialization. We exploit an effective hit/miss predictor~\cite{Alloy} to guide TicToc to use TIC metadata on likely-hit and TOC metadata on likely-miss. The common result: a hit is serviced in one cache access (TIC path), a miss with clean eviction directly goes to memory (TOC path), and a miss with dirty eviction goes to cache and memory (TIC path). An uncommon path of predict-hit actual-miss incurs serialization latency and bandwidth cost to access cache before memory. The other uncommon path of predict-miss actual-hit incurs extra memory access due to parallel lookup of cache and memory.


\subsubsection{TicToc Effectiveness} 

To analyze effectiveness of TicToc, Figure~\ref{fig:tim3} and Figure~\ref{fig:tim4} shows the proportion of channel bandwidth being used for useful operations, install operations, and assorted maintenance operations, for baseline TIC and proposed TicToc. Useful operations include 3D-XPoint Read and Write, and DRAM Cache Hit and Writeback. Install operations refer to cache installs, which are important for improving hit-rate but cost bandwidth to write the line to DRAM. Lastly, Maintenance operations refer to bandwidth-wasting operations used to confirm a line is not resident: miss probes for TIC, and accessing and updating TOC metadata for TOC.

\begin{figure}[htb] 
	\centering
	\vspace{-0.10 in}
    \includegraphics[height = 1.3in]{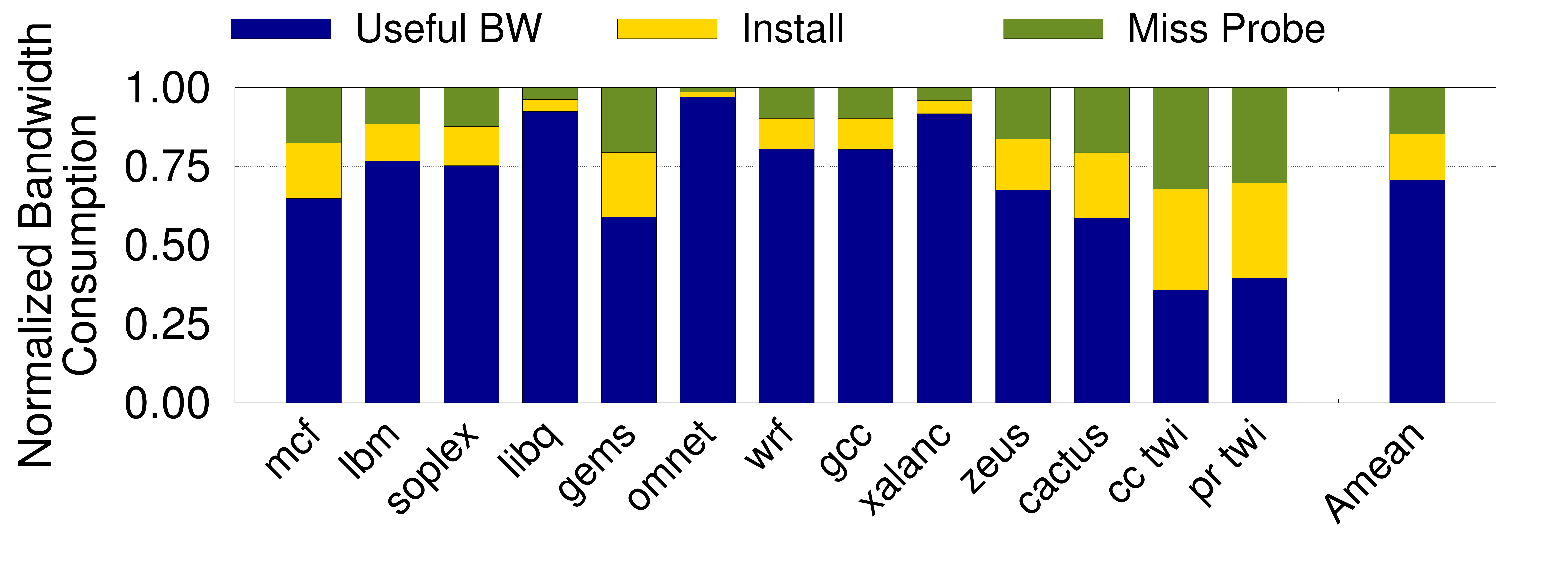}
    \vspace{-0.40 in}
	\caption{Breakdown of bus bandwidth consumption for TIC organization~\cite{Alloy}. Workloads with low hit-rate waste significant bandwidth to confirm misses.}
     \vspace{-0.1 in}
	\label{fig:tim3} 
\end{figure}


As expected, Figure~\ref{fig:tim3} shows that TIC wastes bandwidth probing the DRAM cache to confirm misses. The proposed TicToc can utilize TOC to reduce such miss probes. However, Figure~\ref{fig:tim4} shows that TicToc actually fares worse due to needing bandwidth to maintain TOC tag and TOC dirty-bit. 

TOC tag-updates happen when the workload misses on a line and installs it. A large fraction of misses occur when a workload is accessing many lines in a new page, so misses generally have good spatial locality. In such cases, metadata-accesses/updates are amortized with the small metadata cache. 

TOC dirty-bit-updates, on the other hand, occur upon eviction of a dirty line from an earlier level of cache.
Eviction generally has poor spatial and temporal locality, so updating this information often takes significant bandwidth to read then update the TOC dirty-bit. We need effective methods that target reducing the cost of maintaining dirty information.


\begin{figure}[htb] 
	\centering
	\vspace{-0.10 in}
    \includegraphics[height = 1.3in]{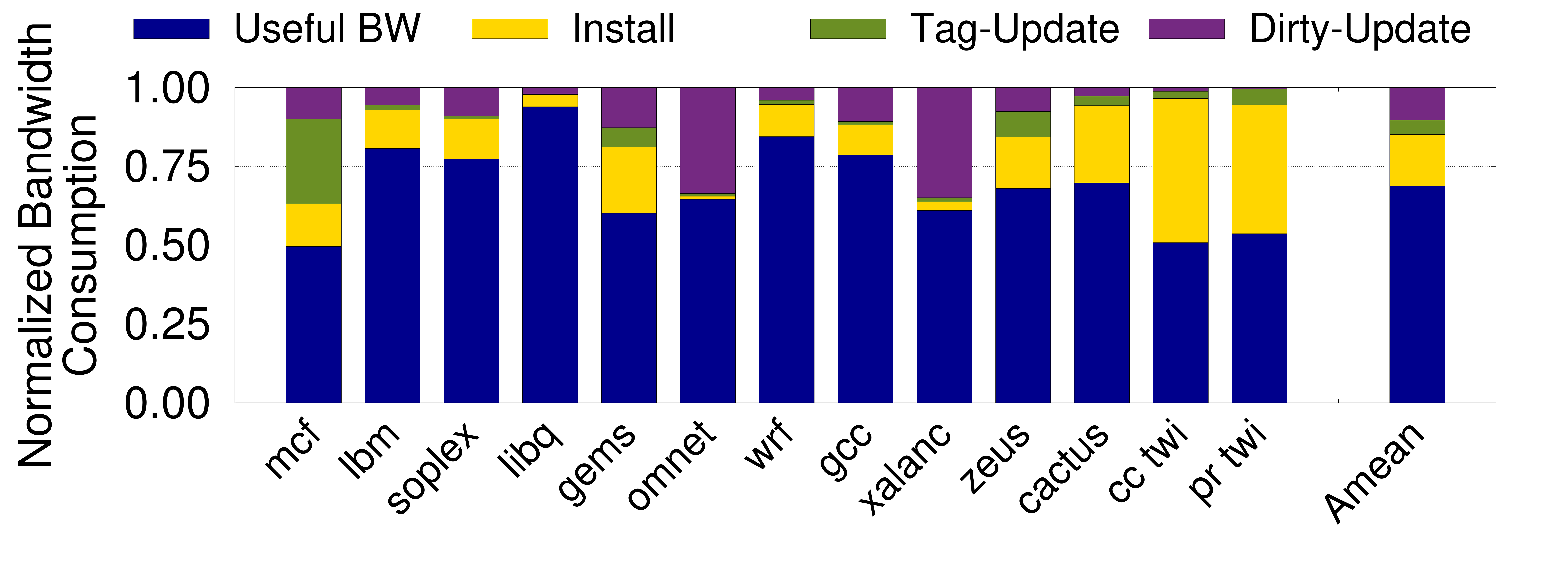}
    \vspace{-0.40 in}
	\caption{Breakdown of bus bandwidth consumption for proposed TicToc organization. Write-heavy workloads waste significant bandwidth updating TOC dirty-bit.}
     \vspace{-0.1 in}
	\label{fig:tim4} 
\end{figure}

\vspace{-0.1 in}
\begin{figure*} 
	\vspace{-0.12in}
	\centering
	\includegraphics[width=\textwidth]{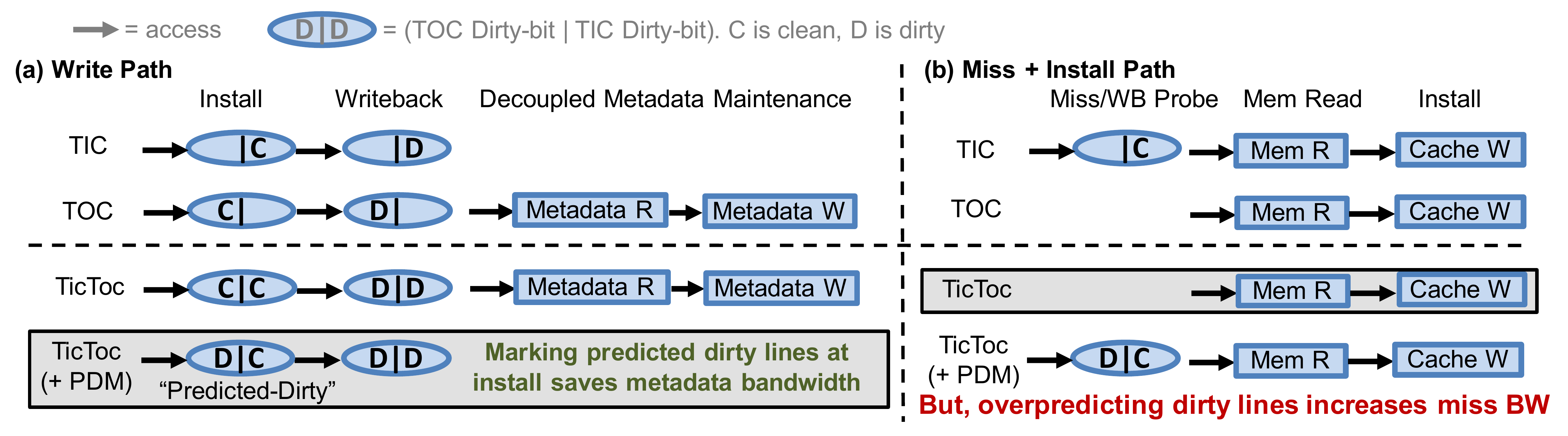}
	\vspace{-0.2in}
	\caption{Bandwidth for a typical (a) write path and (b) miss+install path. TicToc+PDM adds ``Predicted-Dirty'' state, where TOC dirty-bit is installed as dirty but TIC dirty-bit is installed as clean. Installing lines in Pred-Dirty can (a) save TOC dirty-bit update, but (b) increase miss cost. 
	\colorbox{light-gray}{Using Pred-Dirty only for write-likely lines can save bandwidth}
	}
	\label{fig:pdm} 
	\vspace{-0.1in}
\end{figure*}


\subsection{Reducing Dirty-Bit Tracking Costs}

The main source of bandwidth overhead of TicToc is maintaining the dirty-bit for TOC metadata. We need effective methods to reduce the cost of tracking dirty information. We explain difficulty before describing solution.


\subsubsection{Understanding Dirty-bit Updates}

The dirty-bit update procedure starts upon an eviction of a dirty line from L3.  First, we need to check the tag/dirty-bit line in the destination location of the DRAM cache to see if we can overwrite it (i.e., we must first evict a dirty tag-mismatched line). The common case is that the line evicted from L3 is resident in L4 in Figure~\ref{fig:pdm}(a). Chou et al.~\cite{BEAR} proposes to eliminate the tag-check for this common case by maintaining a \textit{DRAM Cache Presence bit (DCP)} alongside every line in L3. The DCP informs us that the same line is in both L3 and L4 -- if the bit is set, the destination location has the same tag and can be directly overwritten (note that this optimizationt is included in our baseline).
Second, the DRAM cache will then write the dirty line to the DRAM cache. Third, the cache will need to update any pertinent tag and dirty-bit metadata. The tag-update for TIC and TOC is uncommon, as typically L3 to L4 writebacks will hit.
The dirty-bit-update for TIC is sent along with L4 install, so it does not incur bandwidth overhead. However, Figure~\ref{fig:pdm}(a)[TOC,TicToc] shows that the dirty-bit update for TOC often needs to be separately queried and potentially updated.  This TOC dirty-bit update is TicToc's main source of bandwidth overhead.

The overhead of dirty-bit updates is comprised of two parts: repeated TOC dirty-bit checks for already-dirty lines, and the initial TOC dirty-bit update to mark clean-to-dirty transition. We target these two scenarios with two techniques.

 


\subsubsection{Reducing Repeated TOC Dirty-bit Checks}

We have an insight that if we also knew the dirty state of the corresponding line in the DRAM cache, we can avoid the need to check the TOC dirty-bit. Instead, we can check (and update) the TOC dirty-bit only if the dirty status changes.

\vspace{.05in}
\noindent \textbf{DRAM Cache Dirtiness:} To enable this optimization, we propose to additionally store a \textit{DRAM Cache Dirtiness bit (DCD)} alongside the DCP~\cite{BEAR} next to each line in the L3 cache. The DCP stores information that the current L3 line is also resident in L4. Meanwhile, the DCD will additionally store the dirty status of that L4 line. We set the DCD on read of a dirty line from L4. On a DRAM cache write, we check both the DCP and DCD. If both DCD and DCP are set, we know the line is resident and already dirty in the TOC metadata -- tag and dirty-bit will be unchanged and we do not need to fetch TOC. Hence, DCP reduces tag checks when tag will not be modified, and DCD reduces dirty-bit checks when dirty-bit will not be modified.

\begin{figure*} 
	\vspace{-0.1in}
	\centering
	\includegraphics[height=1.5in]{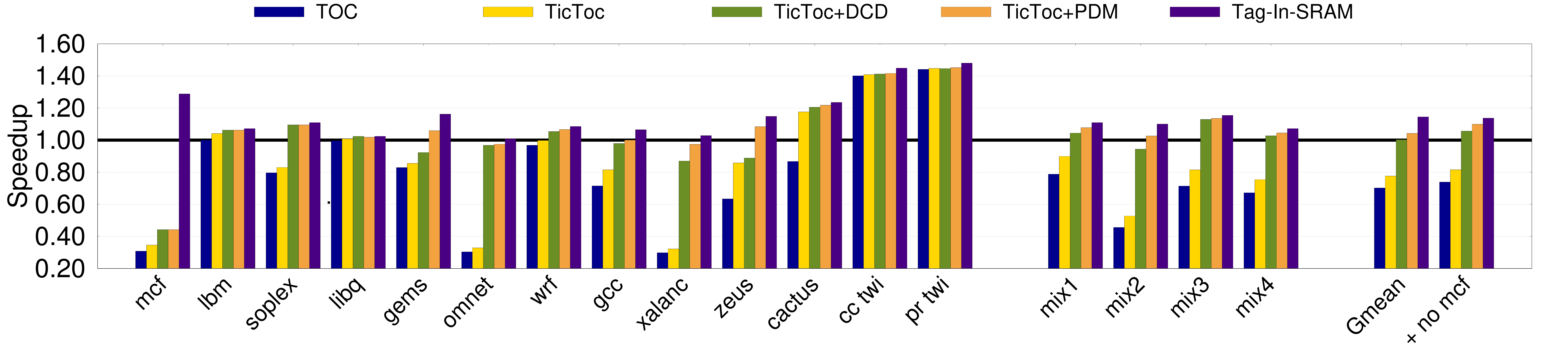}
	\vspace{-0.12in}
	\caption{Speedup of TOC, proposed TicToc, TicToc with DRAM Cache Dirtiness bit, TicToc with Preemptive Dirty Marking (PDM), and ideal Tag-In-SRAM, normalized to TIC. TicToc+PDM performs near ideal for most workloads.}
	\label{fig:perf} 
	\vspace{-0.05in}
\end{figure*}

Figure~\ref{fig:perf} shows that DCD reduces dirty-bit check of many workloads that repeatedly write to same lines (e.g., \textit{omnet}, \textit{soplex}).
However, there are several workloads 
(e.g., \textit{zeusmp}) 
that are write-heavy and write to most lines only once -- we want to reduce dirty-bit updates for those workloads as well.

\subsubsection{Reducing Initial TOC Dirty-bit Update}

For workloads that write-once to lines, we have an insight that if we can preemptively mark the dirty bit in the TOC at install time, we can avoid even the initial TOC clean-to-dirty update that would have occurred at L3 eviction time. 
We call this approach \textit{Preemptive Dirty Marking (PDM)}.

\vspace{.05in}
\noindent \textbf{Preemptive Dirty Marking:} Figure~\ref{fig:pdm} shows the typical write and miss+install bandwidth for TicToc and one that preemptively marks TOC dirty-bit. Figure~\ref{fig:pdm}(a)[TicToc] shows a typical write path needs 4 accesses: a normal line would incur clean install, a write, and TOC dirty-bit read and write. 
Figure~\ref{fig:pdm}(a)[TicToc+PDM] shows that PDM can limit writes to 2 accesses. We add a new dirty state of ``Predicted-Dirty,'' where TOC dirty-bit is marked as dirty but TIC dirty-bit is marked as clean. If we install lines in ``Predicted-Dirty,'' the TOC dirty-bit is set at install time, and even the initial TOC clean-to-dirty update can be avoided.

However, while early marking can save bandwidth on writes, PDM incurs a different problem on the miss path. Figure~\ref{fig:pdm}(b)[TicToc] shows a typical miss+install path needs 2 accesses: TOC metadata informs residence and dirtiness so miss+install can be accomplished with a memory read and a DRAM cache install. However, Figure~\ref{fig:pdm}(b)[TicToc+PDM] shows that PDM can increase miss+install to 3 accesses. For instance, if an otherwise clean line has been preemptively marked as dirty in the TOC dirty-bit, we would read the DRAM cache line in preparation for an eviction of a dirty line, thereby adding an extra DRAM read. Note that the Predicted-Dirty state does not cause extra memory writebacks as the miss/wb probe will find the TIC dirty-bit, and write back only if the data is dirty. Thus, being aggressive in marking lines as ``Predicted-Dirty'' will save write bandwidth, but it can come at the cost of increasing miss bandwidth.

Ideally, we want to avoid write costs by installing write-likely lines as ``Predicted-Dirty'', and avoid increased miss costs by installing write-unlikely lines as clean.  However, if we install a write-likely line as clean, it will pay increased miss cost. Conversely, if we install a write-unlikely line as ``Predicted-Dirty,'' it will pay cost to update TOC dirty bit. Hence, performance of Preemptive Dirty Marking is contingent on good classification of write-likely and write-unlikely lines at install-time to avoid both TOC dirty bit update and TIC miss probe bandwidth.

\begin{figure}[htb] 
	\centering
    \includegraphics[height = 1.6in]{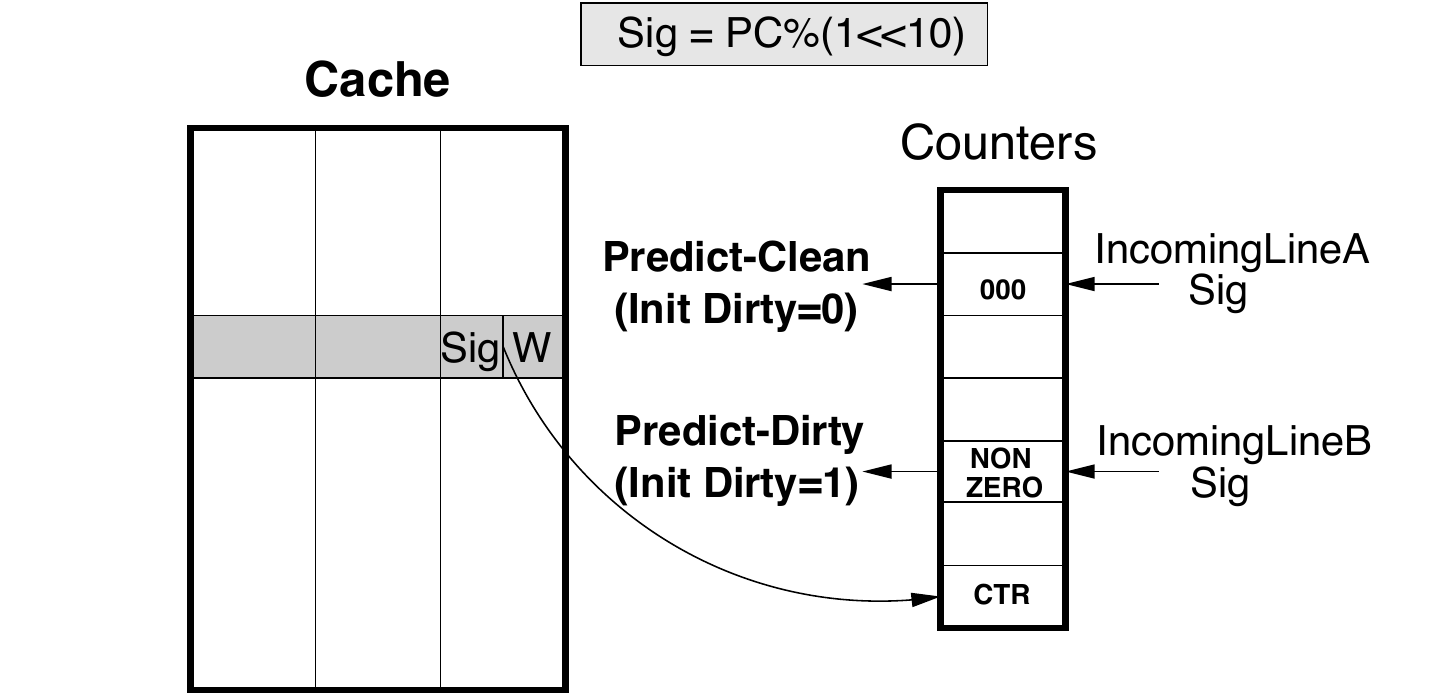}
    \vspace{-0.10 in}
	\caption{Signature-based Write Predictor learns which sigs correspond to eventual write, to aid PDM technique.}
     \vspace{-0.1 in}
	\label{fig:writepred} 
\end{figure}

\vspace{.05in}
\noindent \textbf{Write Predictor:} For accurate write-classification for PDM, we develop a \textit{Signature-based Write Predictor (SWP)} to predict likeliness an incoming line will be written. SWP employs a sampling PC-based prediction, inspired by SHiP~\cite{SHIP,SHIP++}. 
Figure~\ref{fig:writepred} shows structures and operation of SWP.
SWP consists of write-behavior observation, learning, and prediction.

Observation is accomplished by maintaining signature (installing-PC in this case) and a written-to bit inside the metadata of each line (10 bits additional metadata for the 1\% sampled lines, stored in TOC-metadata). Signature is set at install-time, and written-to bit is updated on first write to line. On eviction of such a sampled line, we get the information that this PC installed a line that was either written-to or never written-to in its lifetime in the cache.

Learning is then accomplished by storing observed write-behavior into a PC-indexed table of saturating 3-bit counters. On eviction of a line that has the written-to bit set, the counter corresponding to installing-PC is incremented. On eviction of a line that does not have written-to bit set, the counter corresponding to installing-PC is decremented. This counter table becomes a PC-indexed table of write-behavior.

Prediction is then simple -- on install, the installing-PC is used to index into the counter-table to provide a write-likely or write-unlikely prediction. If the counter is non-zero, this PC has seen write behavior and the incoming line should be installed in ``Predicted-Dirty'' state to avoid initial TOC clean-to-dirty update. If the counter is zero, then this PC has not seen much write behavior and the incoming line should be installed as clean to avoid miss/wb probes.

\begin{figure}[htb] 
	\centering
	\vspace{-0.05 in}
    \includegraphics[height = 1.3in]{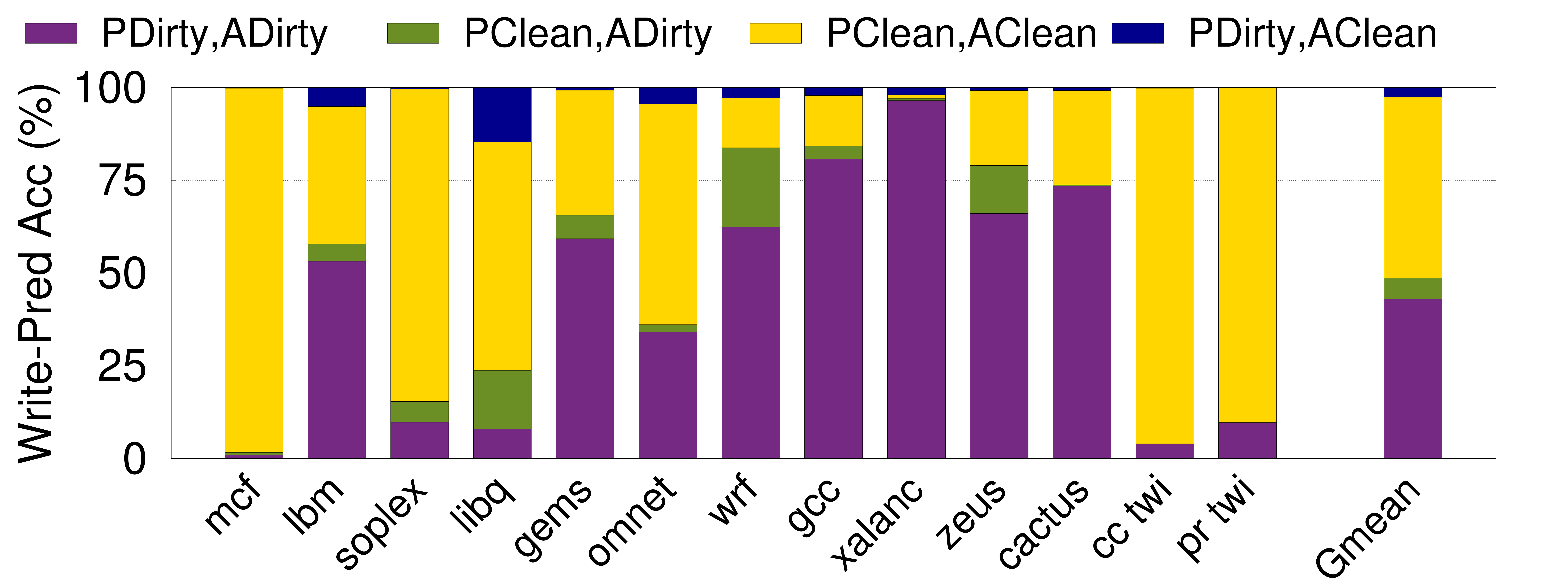}
    \vspace{-0.3 in}
	\caption{Accuracy of Write Prediction (P=predicted, A=actual). Low PClean/ADirty and PDirty/AClean rate reflects accurate write-behavior prediction.}
     \vspace{-0.12 in}
	\label{fig:writepred_acc} 
\end{figure}

\vspace{.05in}
\noindent \textbf{Accuracy of Write Predictor:}
Effectiveness of PDM is contingent on good classification of write-likely (dirty) lines to reduce dirty-bit update cost, and write-unlikely (clean) lines to reduce miss-probe cost. Figure~\ref{fig:writepred_acc} shows the fraction of lines that are predicted clean or dirty, and actually clean or dirty.
On average, SWP predicts clean and dirty with 92\% accuracy, and enables PDM to save most dirty-update and miss-probe bandwidth.

\begin{figure*}[htb] 
	\centering
    \includegraphics[height = 1.7in]{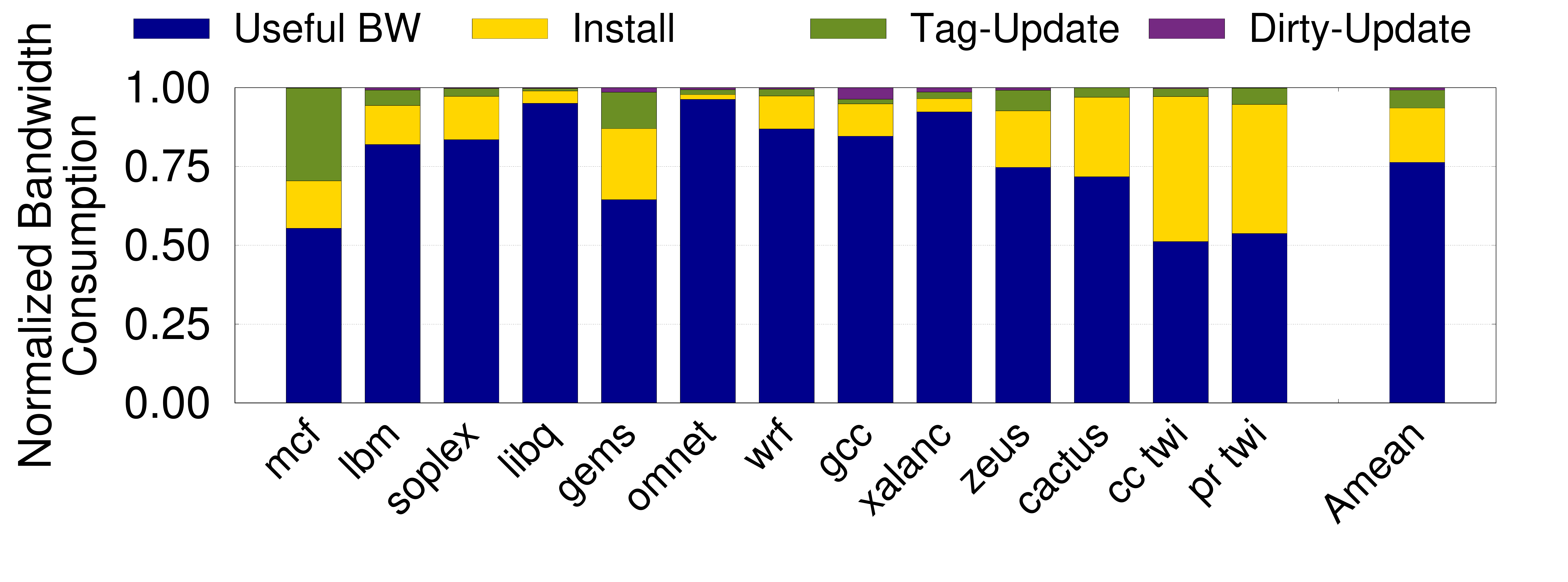}
    \vspace{-0.20 in}
	\caption{Breakdown of bus bandwidth for dirty-optimized TicToc. Dirty-bit updates are greatly reduced.} 
	\label{fig:tim_ta} 
\end{figure*}

\begin{figure*}[htb] 
	\centering
    \includegraphics[height = 1.7in]{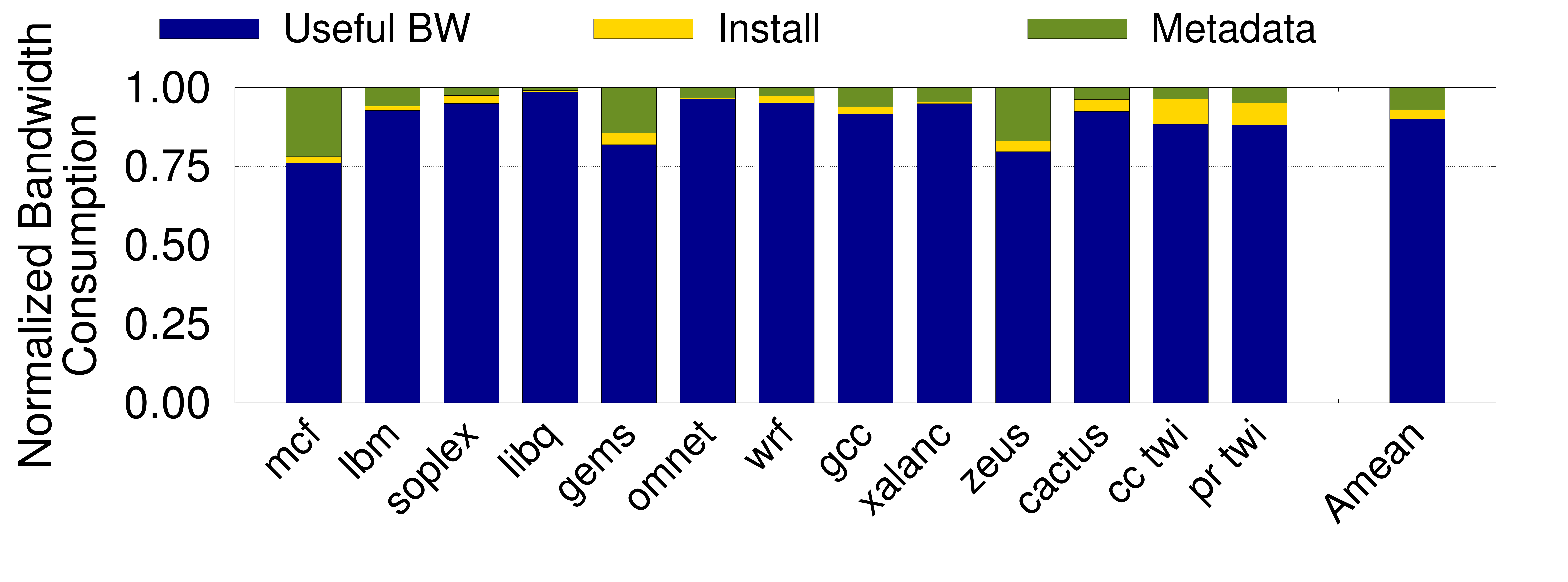}
    \vspace{-0.20 in}
	\caption{Breakdown of bus bandwidth for dirty-optimized TicToc w/ Write-Aware Bypassing. Installs are mitigated.}
    \vspace{-0.03 in}
	\label{fig:tim_tab} 
\end{figure*}

\subsubsection{Effectiveness of Dirty-Tracking Optimizations}

\textbf{Performance:} Figure~\ref{fig:perf} shows the speedup of TOC, our TicToc, TicToc with DCD, TicToc with PDM, and idealized Tag-In-SRAM, normalized to TIC approach. TOC performs poorly due to poor metadata-cache hit-rate, for 30\% slowdown. Our TicToc reduces hit bandwidth, for 22\% slowdown. Adding DRAM Cache Dirtiness bit reduces dirty-bit tracking for repeated writes to same lines, for 0\% speedup. Adding Preemptive Dirty Marking reduces the initial dirty-bit update without incurring extra miss bandwidth due to accurate Write Predictor. Notably, TicToc+PDM achieves near idealized Tag-In-SRAM performance for most workloads for 10\% speedup, without including worst-case \textit{mcf} in the average. Few workloads see performance gap to ideal. We analyze bandwidth consumption to gain insight into the problem.



\textbf{Bandwidth:} Figure~\ref{fig:tim_ta} shows the bandwidth breakdown of TicToc + dirty-bit optimizations. Overall, our approach eliminates nearly all of the TOC dirty-bit update bandwidth (decreased fraction from 10\% to 0.8\%) and frees up bandwidth for useful reads and writes.
However, we note that installing lines and updating the TOC-tag now becomes the primary source of DRAM cache bandwidth overhead. We target this overhead next.


\section{Reducing Install Bandwidth \\ With Write-Aware Bypass}

When data has poor reuse, installing lines and updating TOC metadata wastes bandwidth. In fact, in such cases, employing a DRAM cache could actually hurt performance, as the line install and tag maintenance operations needlessly steal bus bandwidth from memory accesses.
Figure~\ref{fig:perf_bypass} shows the performance of a setup without a DRAM cache,
normalized to a setup with a TIC DRAM cache. We note that there are multiple workloads (e.g., \textit{pr twi} and \textit{cc twi}), for which ``no DRAM cache'' performs better than TIC. 
While one can avoid this degradation by disabling the DRAM cache at boot-time, doing so would then hurt the cache-friendly workloads. Therefore, we need effective mechanisms to reduce the cost of unnecessary installs. 

\vspace{.15in}
\noindent \textbf{Insight -- Write-Aware Bypassing:} Prior work has proposed cache bypassing~\cite{BEAR,ctrbypass,crc1} to avoid unnecessary installs. On an L3 miss, one can bypass the DRAM cache and install the line only in L1/L2/L3 caches, thereby saving the DRAM cache install bandwidth. However, such bypassing must be done selectively and carefully, otherwise it may increase writes to 3D-XPoint memory, and degrade performance, endurance, and power.

\newpage

\subsection{Design of Write-Aware Bypassing}
\begin{figure}[htb] 
	\centering
	\vspace{-0.12 in}
    \includegraphics[width=\columnwidth]{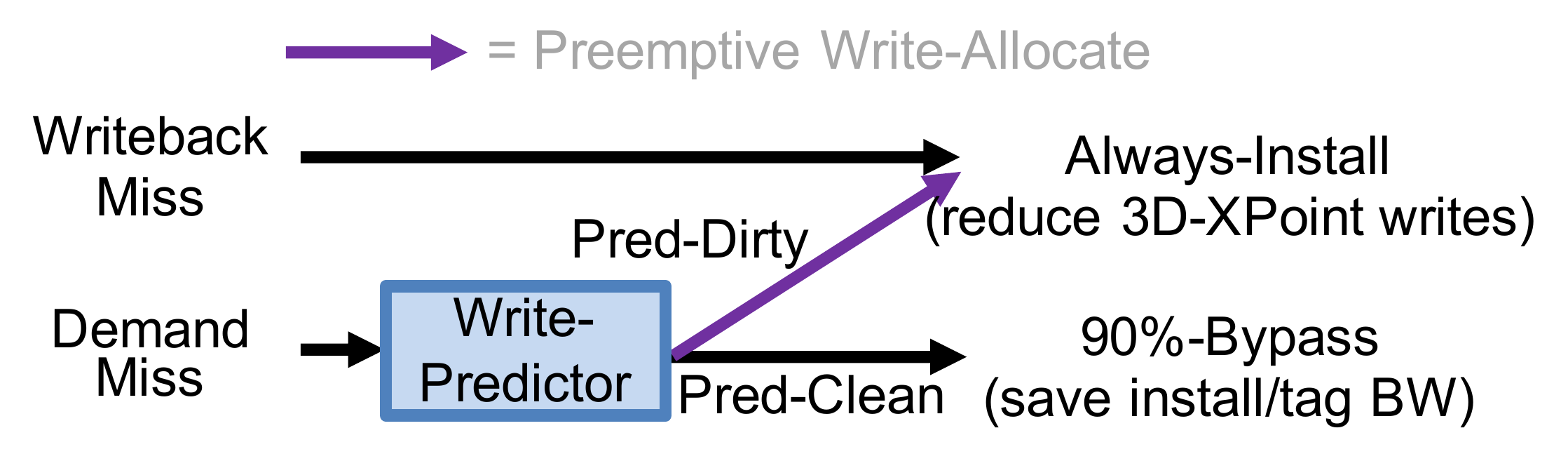}
    \vspace{-0.30 in}
	\caption{Write-Aware Bypass. Reduce install bandwidth by bypassing most write-unlikely lines. Reduce 3D-XPoint writes by installing write-likely lines.}
     \vspace{-0.18 in}
	\label{fig:bypass} 
\end{figure}

\begin{figure*}[htb] 
	\vspace{-0.3in}
	\centering
	\includegraphics[height=1.5in]{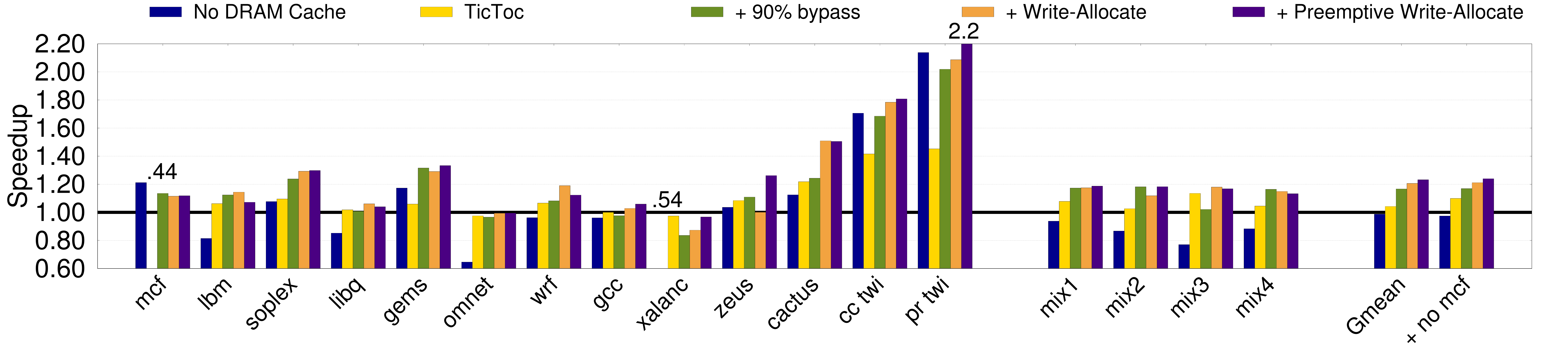}
	\vspace{-0.18in}
	\caption{Speedup of a no-DRAM-cache configuration, proposed TicToc organization, adding 90\%-bypass, adding Write-Allocate, and adding Preemptive Write-Allocate, relative to TIC approach.} 
	\label{fig:perf_bypass} 
	\vspace{-0.08in}
\end{figure*}

Figure~\ref{fig:bypass} shows our Write-Aware Bypassing policy.
We start with the default 90\%-bypass policy proposed in~\cite{BEAR}, which bypasses 90\% of all installs. While such aggressive bypassing was shown to work well for an HBM+DDR hybrid memory~\cite{BEAR}, we note that it can increase write traffic to the write-constrained 3D-XPoint memory. To address this problem, we add write awareness to the bypass policy. We augment the default bypass policy with a write-allocate condition, which requires that dirty L3 evictions would \textit{always} install DRAM cache lines. Thus, the DRAM cache would act as a write buffer for 3D-XPoint memory. Unfortunately, the drawback of such an approach is that installing DRAM cache lines at the time of L3 evictions may result in significant tag-update costs.
L3 evictions often have poor spatial locality, so
TOC tag updates carried out at L3 eviction time exhibit poor metadata cache hit rates and incur extra DRAM accesses. 

To amortize the TOC tag-update cost of our write-allocate policy, we propose \textit{Preemptive Write-Allocate}, whereby we also \textit{always-install} write-likely lines (predicted with SWP). Preemptive Write-Allocate enables our write-allocate installs to happen at L3 miss time. Such installs have higher spatial locality, resulting in more metadata cache hits and more effective amortization of TOC metadata updates.

\vspace{-.1in}

\newpage

\subsection{Effectiveness of Write-Aware Bypassing}

\noindent \textbf{Bandwidth:}
To understand the effectiveness of our install and metadata-update reducing optimizations, we show the bandwidth breakdown of our approach in Figure~\ref{fig:tim_tab}. Overall, we find that install-reducing optimizations can eliminate nearly all of the install bandwidth overheads and leave much more bandwidth for useful reads and writes. In total, the combination of our cache bandwidth reducing optimizations improves fraction of bandwidth going to useful operations (servicing reads / writes) from 70\% to 90\% on average.

\vspace{.05in}
\noindent \textbf{Performance:}
Figure~\ref{fig:perf_bypass} shows the performance of TicToc with dirty-optimizations, TicToc with 90\%-bypass, TicToc with 90\%-bypass and write-allocate, and TicToc with 90\%-bypass and preemptive write-allocate, relative to TIC. 

TicToc with dirty-bit optimizations does well for most workloads for an average 4.2\% speedup, but can suffer for workloads with poor spatial locality and low hit-rate (e.g., \textit{mcf}). TicToc with 90\%-bypass reduces install and TOC tag-update cost to improve speedup to 16.7\%. Notably, the performance degradation for \textit{mcf} has been substantially mitigated. TicToc with 90\%-bypass and write-allocate enables effective write-buffering to improve speedup to 20.6\%. Finally, TicToc with 90\%-bypass and preemptive write-allocate further amortizes TOC metadata-update (e.g., useful for \textit{zeusmp} and \textit{pr twi}) to improve speedup to 23.2\%. 

\subsection{Putting it all together}

Overall, our proposed techniques target all forms of DRAM cache maintenance bandwidth to achieve a bandwidth-efficient (\textgreater 90\% of channel bandwidth to useful operations) and low SRAM storage overhead (34KB) DRAM cache organization: \textit{TicToc} improves hit and miss bandwidth, \textit{DRAM Cache Dirtiness bit} and \textit{Preemptive Dirty Marking} reduces dirty-bit-tracking bandwidth, and \textit{Write-Aware Bypass} reduces install and tag-tracking bandwidth.
Our TicToc with dirty-bit and install bandwidth reducing optimizations enables 23.2\% speedup at the cost of only 34KB of SRAM.



\ignore{

        TODO TODO
    
        8GB DRAM channel. 3DXP channel
        4GB DRAM + 3DXP channel. 4GB DRAM + 3DXP channel.

    With Prefetching. 
        Prior works commonly use prefetching to gain back performance. But obfuscates pure bandwidth costs of prior proposals. 

Prefetcher on our TIMALLOY    
    NL, Double, Quad, Aggressive (hybrid stride+delta prefetcher, tuned version of [slim ampm]).
        Tuning
            Prefetch on access stream (both hit/miss). Better patterns.
                Prefetch filtering: normally, have an already-pf'ed bit to avoid prefetching again on re-access of resident lines. But, updating bit requires a separate write. Instead, Prefetching turned on only after first miss to page. Approximates already-pf'ed bit.
            Double/Quad need to modify hit/miss predictor as well.
            Prefetch buffer to avoid repeated probes to already-accessed lines.
            If bypass chosen for prefetched line, forwarded to L3 cache.
Prefetcher on Alloy (high repeated miss probe cost), Timber (hit bw, high update-costs).
       
        Assumed direct-mapped so 1 access gets line (alloy). Associativity sometimes causes double-accesses to find line, and wastes bandwidth. Can use ACCORD. However, multiple accesses to find line is usually not a good tradeoff as it comes at cost to bus bandwidth of memory.

Timber sizing. write improvement improves scaling
        More graceful perf curve to smaller sizing
        if TIMBER sizing too small to capture 
        128/256/512/1024 ta5\_preddirty
        128/256/512/1024 ta3 TA

}

\section{Results and Discussion}
In this section we present sensitivity studies and storage analysis. Due to space constraints, we limit results to TicToc with dirty-bit optimizations.



\subsection{Storage Requirements}

We analyze the SRAM storage requirements of our TicToc organization. TicToc requires structures from its component TIC and TOC organizations. Inheriting from TIC, we need \textasciitilde 1KB for PC-based hit/miss prediction~\cite{Alloy}, and 1 bit alongside each L3 line for DRAM Cache Presence bit to avoid tag-check for writes to resident lines~\cite{BEAR}. Inheriting from TOC, we need 32KB for a metadata cache~\cite{timber}. 

Specific to TicToc, to implement our dirty-bit optimizations, we need a 1-bit bit alongside each L3 line for DRAM Cache Dirtiness, and \textasciitilde 1KB for our Signature-based Write-Predictor (512 entries of 3-bit counters with 9-bit PC tag). Our bypassing optimizations do not require additional space. 
In total, TicToc needs 34KB SRAM storage in the memory controller, with 2 bits alongside each L3 line. 


\begin{table}[hbt]
	\vspace{-.16 in}
	\centering
		\caption{Storage Requirements of TicToc}
		\vspace{.1 in}
		\begin{tabular}{|c||c|}\hline
			TicToc Component & SRAM Storage  \\ \hline \hline
			Hit-Miss Predictor~\cite{Alloy}  & 1 KB \\ \hline
			DRAM Cache Presence~\cite{BEAR}  & 1-bit / L3-line \\ \hline
			Metadata Cache~\cite{timber}  & 32 KB \\ \hline
			DRAM Cache Dirtiness & 1-bit / L3-line  \\ \hline
			Signature-based Write Predictor  & 1 KB  \\ \hline \hline
			TicToc  & 34KB + 2-bits/L3-line \\ \hline
		\end{tabular}
		\label{fig:storage}
	\vspace{-.12 in}
\end{table}







\subsection{Sensitivity to Metadata-Cache Size}

The largest SRAM component of our TicToc proposal is the TOC metadata cache. Table~\ref{tab:size} shows performance sensitivity of our TicToc organization to metadata-cache sizing. We show average speedup of TicToc with dirty-bit optimizations, when employing metadata-caches with sizes ranging from 8KB to 64KB. The dirty-bit tracking optimizations enable TOC approaches to be much more effective with small metadata-cache sizes, as the metadata caches do not need to sized to handle writeback traffic that has poor spatial locality.


\begin{table}[hbt]
	\vspace{-.16 in}
	\centering
	\caption{Sensitivity to Metadata Cache Sizing}
	\vspace{.1 in}

	\begin{tabular}{|c||c|c|}\hline
		Num. Entries  & TicToc  &  TicToc (no mcf) \\ \hline \hline
		128 (8KB)     & -3.0\%          & +1.9\%          \\ \hline
		256 (16KB)    & +1.5\%          & +7.0\%          \\ \hline
		512 (32KB)    & +4.2\%          & \textbf{+10.0\%} \\ \hline
		1024 (64KB)   & +4.8\%          & +10.7\%         \\ \hline
\ignore{
        128 (8KB)   & +15.1\%  \\ \hline
		256 (16KB)  & +19.7\% \\ \hline
		512 (32KB)   &  \textbf{+23.2\%} \\ \hline
		1024 (64KB)   &  +24.5\% \\ \hline
}
	\end{tabular}
	\label{tab:size}
	\vspace{-.12 in}
\end{table}

\subsection{Impact of Channel-Sharing}
\label{ssec:channel_sharing}

\begin{figure}[htb] 
	\vspace{-0.13in}
	\centering
	\includegraphics[width=1.0\columnwidth]{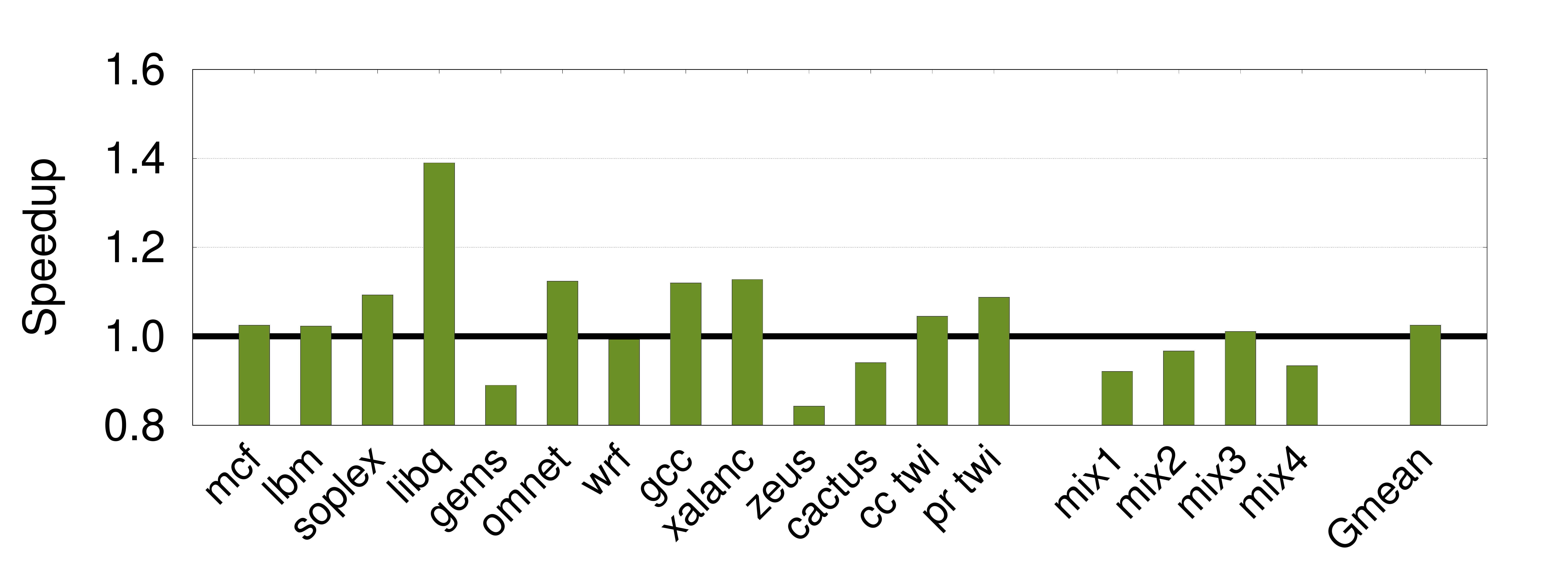}
	\vspace{-0.27in}
	\caption{Speedup of Channel-Shared Hybrid Memory, over Dedicated-Channel Hybrid Memory. Channel-sharing enables up to 40\% speedup.}
	\label{fig:perf_sharing} 
	\vspace{-0.10in}
\end{figure}

DRAM and 3D-XPoint are likely to be behind the same channel to maximize the bandwidth out of each physical pin, as shown in Figure~\ref{fig:intro}.
Figure~\ref{fig:perf_sharing} shows the system performance of a \textit{channel-shared} system (two channels of TIC DRAM cache + 3D-XPoint), normalized over previously assumed \textit{dedicated-channel} systems (one channel of 2x TIC DRAM cache, and one channel of 2x 3D-XPoint).
We find that channel-shared systems enable more balanced channel bandwidth usage due to each channel having a DRAM cache. For example, under high DRAM cache hit-rate, a channel-shared system would be able to utilize all channels, whereas a dedicated-channel system would only be able to use the half of channels employing DRAM caches. Such channel-shared approaches enable up to 40\% speedup compared to the traditional dedicated-channel setups.

\subsection{Multi-programmed Workloads}
\label{ssec:mixed}

\begin{figure}[htb] 
	\centering
	\includegraphics[width=1.0\columnwidth]{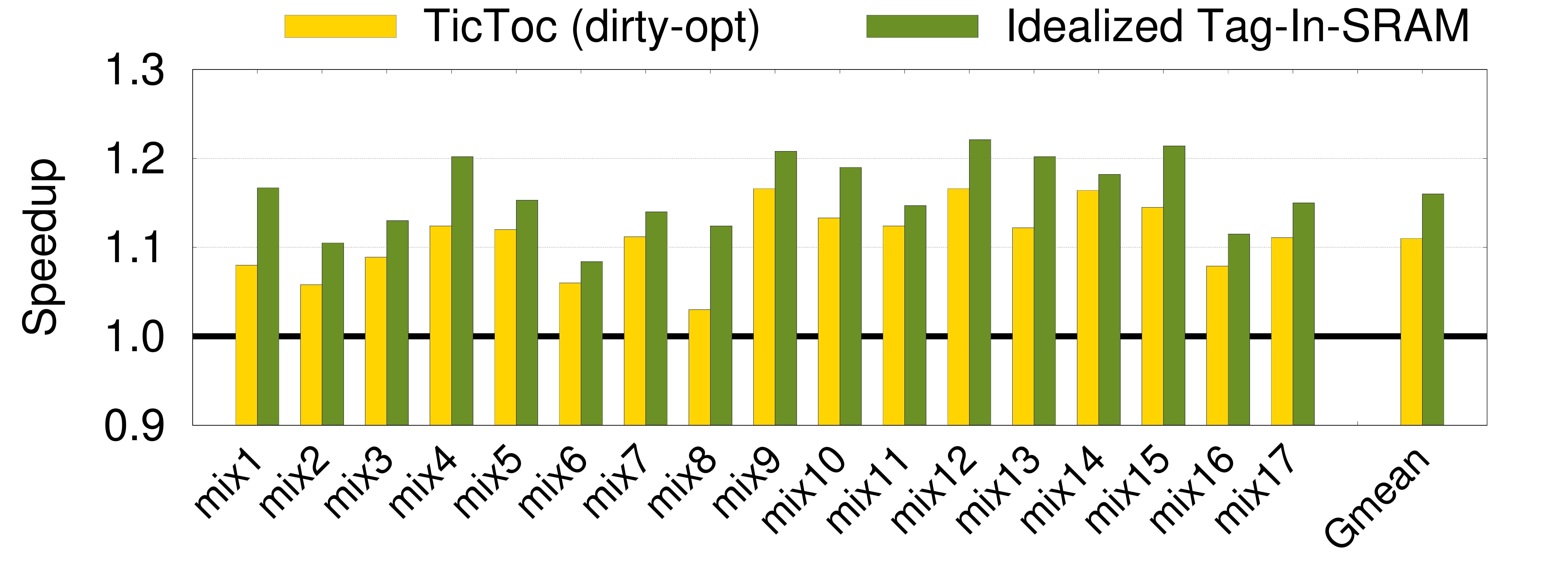}
	\vspace{-0.30in}
	\caption{Speedup of TicToc with dirty-bit optimizations, and idealized Tag-In-SRAM, on mixed workloads.}
	\label{fig:mix_perf} 
	\vspace{-0.13in}
\end{figure}

To show robustness of our proposal to multi-programmed workloads, we conduct evaluations over a larger set of 17 mix-application workloads. 
Figure~\ref{fig:mix_perf} shows that our dirty-optimized TicToc organization provides 11\% speedup across 17 mixes, with no workloads experiencing slowdown.

\subsection{Performance Gap to DRAM-only Solution}

\begin{figure}[htb] 
	\centering
	\includegraphics[width=1.0\columnwidth]{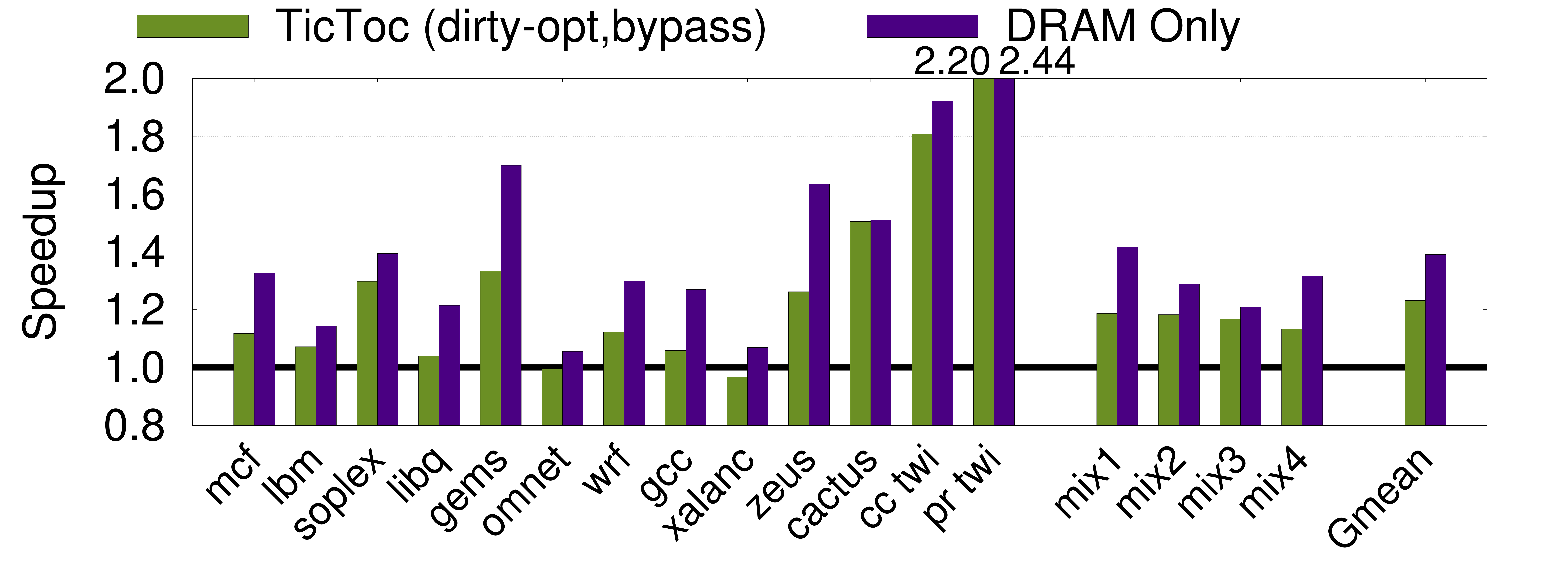}
	\vspace{-0.23in}
	\caption{Speedup of TicToc (dirty-opt, bypassing) and DRAM-only solution, relative to TIC cache + 3D-XPoint.}
	\label{fig:perf_dram} 
	\vspace{-0.10in}
\end{figure}

In order to quantify the remaining performance opportunity, we compare our TicToc DRAM cache + 3D-XPoint solution with an expensive DRAM-only solution having the same DRAM main memory capacity as the 3D-XPoint capacity in our setup. Note that this DRAM-only solution will cost substantially (4--8x) more than a hybrid DRAM+3D-XPoint memory. Figure~\ref{fig:perf_dram} shows performance results normalized to TIC. TicToc's bandwidth-efficient DRAM caching enables 3D-XPoint to perform within 13\% of the expensive DRAM-only solution.

\subsection{Considerations for Associativity}



Our TicToc implementation uses a direct-mapped organization to avoid the latency of serialized tag and data lookup (TOC explanation in Section~\ref{ssec:our_TOC}). There have been works that can avoid the serialized tag lookup for associative TIC~\cite{ACCORD} and associative TOC~\cite{unison} designs via scalable way-prediction methods. As such, we find associativity an orthogonal issue for our proposal; techniques such as~\cite{ACCORD} can be incorporated into our design.


\ignore{

TicToc targets improving bandwidth-efficiency of DRAM cache designs. It assumes a baseline direct-mapped cache similar to commercial design~\cite{KNL}.
TicToc targets bandwidth-efficiency

}



\ignore{
\subsection{Impact of Prefetching}

Data Prefetching improves memory latency at the cost of bandwidth.
Prior approaches have incorporated prefetching into their DRAM caches and add it into their cache performance benefits. However, we find prefetching a largely orthogonal enhancement, as prefetching has a similar path and bandwidth costs to accessing a line that will miss. 

As such, we choose to analyze prefetching separately from DRAM cache organization, and analyze the impact of different prefetchers for different underlying DRAM cache organizations in Figure~\ref{tab:prefetch}. Prefetching has a positive effect on performance as expected, but we find that prefetching cannot fundamentally hide the bandwidth overheads of underlying cache designs. Prefetching on top of our proposed bypassing dirty-optimized Dual-Tag approach improves speedup from 22.4\% to 26.3\% speedup relative to baseline Distributed-Tag approach. This is close to the upper-bound of 36.9\% speedup possible when using an ideal DRAM-only configuration.

\begin{table}[hbt]
	\vspace{-.21 in}
	\centering
	\caption{Prefetch on various DRAM Caches}
	\vspace{.03 in}

\begin{tabular}{|l|c|c|c|}
\hline
     & Distributed & Aggregated & Proposed Dual-Tag \\ \hline \hline
Base     & +0.0\%      & -13.5\%    & \textbf{+22.4\%}                            \\ \hline
Nextline & +3.9\%      & -13.4\%    & +24.4\%                            \\ \hline
Quad     & +7.6\%      & -12.4\%    & +26.3\%                            \\ \hline
Stride   & +4.2\%      & -11.7\%    & +26.0\%                            \\ \hline
\end{tabular}
	\label{tab:prefetch}
	\vspace{-.15 in}
\end{table}

}


\ignore{
    Related
      Line-based
        BEAR
            BEAR solves writeback probe with DCP bit. use this.
            BEAR's miss probe reduction not applicable to current systems
            BEAR's bypassing is too coarse-grain.
        TIMBER
      Page-based      
        Unison
            todo. 
            
      Mostly-clean for effective miss speculation. Also follow-up for f

      DRAM Cache for PCM. Highly associative, poor hit latency. 
      
      With OS support, tagless/banshee. get capacity. page granularity movement. Swaps increase writes.

}

\section{Related Work}


\subsection{Line-based DRAM Caches}

In our work, we utilize and combine the two major types of line-granularity DRAM cache designs: \textit{Tag-Inside-Cacheline (TIC)} and \textit{Tag-Outside-Cacheline (TOC)} approaches. 

\textbf{TIC} designs~\cite{Alloy,BEAR,candy,ACCORD,DICE,KNL} organize their cache as direct-mapped and store tag inside the cacheline, such that one access can retrieve both tag and data. Such approaches are optimized for hits, but pay bandwidth to confirm misses~\cite{Alloy}. BEAR~\cite{BEAR} proposes several enhancements to reduce bandwidth cost of cache maintenance: we include its DRAM Cache Presence that targets reducing write probe in our baseline TIC design, we compare with Bandwidth-Aware Bypass with 90\%-bypass in Figure~\ref{fig:perf_bypass}, but, however, we do not include Neighboring Tag Cache as current implementations cannot obtain neighboring tag for free~\cite{KNL}. 
Such hit-latency optimized approaches have been proven effective in industrial application with Intel Knights Landing product~\cite{KNL}; as such, we perform all of our experiments with BEAR as our baseline.
We use BEAR as our TIC component of TicToc, and improve upon TIC miss-bandwidth inefficiencies to enable a scalable bandwidth-efficient DRAM cache.



\textbf{TOC} designs~\cite{LHCache,atcache,bwp,simcache,timber} store tags in a separate area of the DRAM cache and fetch them as needed. The earliest forms of such caches were highly associative and would need a serial tag then data lookup~\cite{LHCache}. Some enhancements used tag-prefetching~\cite{atcache} or way-prediction~\cite{bwp} to avoid this serialized tag lookup. Others used direct-mapped organization~\cite{simcache,timber} to avoid serialized tag lookup, with one employing a tag cache~\cite{timber} to reduce the bandwidth of tag lookup as well. Figure~\ref{fig:perf} shows that TIMBER~\cite{timber}, a direct-mapped TOC design with tag cache, performs well for misses but can perform poorly due to high bandwidth cost to update metadata. We use TIMBER as our TOC component of TicToc, and improve upon TOC metadata-bandwidth inefficiencies to enable a scalable bandwidth-efficient DRAM cache.

\subsection{Page-based DRAM Caches}

An alternate approach to designing DRAM-caches is to use large-granularity caches to amortize tag and metadata overhead, in hardware~\cite{FootprintCache:ISCA2013:MICRO2014papers,unison} or software~\cite{tagless,tagless2,loh_freq,banshee}.

\vspace{0.1 in}
\textbf{Hardware-only:} Hardware-based approaches store tags either in SRAM~\cite{FootprintCache:ISCA2013:MICRO2014papers,sector} or in DRAM~\cite{unison}. The Tag-In-SRAM proposals typically use sector caching~\cite{sector} to reduce the overall tag requirements, and fit them all in MegaBytes of SRAM~\cite{FootprintCache:ISCA2013:MICRO2014papers}. However, the storage for these approaches are still typically quite large. And, they have the penalty of poor cache utilization when not all lines in a page are used. For comparison, we show what these cache organizations can achieve with the line-granularity Tag-In-SRAM organization in Figure~\ref{fig:perf}. Our proposed TicToc achieves close to this upper-bound with much less SRAM storage (34KB vs. \textgreater 20MB).

Alternatively, there are Tag-In-DRAM proposals that store metadata in DRAM, and fetch the tags as needed~\cite{unison}. These approaches need to spend bandwidth to access and update metadata information in a separate area of DRAM cache. As such, these approaches often have similar bandwidth overheads and performance to the TOC component of TicToc. And, they have the penalty of poor cache utilization when not all lines in a page are used. For comparison, we show what these cache organizations can achieve with the line-granularity TOC organization in Figure~\ref{fig:perf}. Our proposed TicToc, as well as the baseline TIC, outperforms such TOC organizations.
Nonetheless, our dirty-tracking optimizations are general and can be applied to improve metadata update cost for such caches as well.

\textbf{Software-supported:} Software-supported DRAM cache approaches maintain mapping and metadata information inside page tables~\cite{tagless,tagless2,loh_freq,banshee}, and use various heuristics to determine when to install pages. The benefit to such approaches is that they do not need to pay additional bandwidth to access tags.
The shortcomings of such an approach are two-fold. First, the migration granularity is fixed to the size of OS page, which can cause overfetch problems, as well as poor cache DRAM utilization when not all of the page is useful.
Second, such approaches require both hardware and software support, and can be difficult to deploy without cooperation between multiple vendors. We do not perform comparison with such works as these approaches are out of scope (i.e., break our design goal of OS-transparency).

\subsection{Two Level Memories}

Other hybrid memory approaches attempt to get the capacity of both memories, and instead initiate hardware-managed line or page \textit{swaps} to enable most data to be serviced at the lower-latency or higher-bandwidth memory~\cite{Chou:micro2014,Sim:micro2014,silc-fm,mempod,pageseer}. These approaches have various tracking overheads and effectiveness. However, we note there is a fundamental difference from caching. On eviction of an unmodified line/page, caches can simply drop the clean line/page -- whereas, swap-based approaches need to always write back the evicted line/page. Such swaps incur extra writes that could otherwise have been avoided. For our target DRAM + 3D-XPoint configuration, these extra swapping-induced writes would cost performance, endurance, and power when writing to write-constrained 3D-XPoint. 
The added capacity benefits (3-12\%) obtained from such swapping are unlikely to make up the difference. Hence, we do not take a swapping-based / two-level memory approach for this work. We do not perform comparison with such works as these approaches are out of scope (i.e., break our design goal of write-efficiency). 



\subsection{On Reducing Dirty-bit Tracking}

Tracking dirty-bit or most-recent-copy of cacheline efficiently with low SRAM storage costs is a known difficult problem. Many works limit the amount of lines that can be kept dirty~\cite{Sim:MICRO2012,c3d}, to reduce SRAM storage costs needed to track dirty lines. Other approaches are more extreme and make the cache clean-only by always writing through~\cite{GPUvi,CARVE}. However, for our work, we target a DRAM + 3D-XPoint system, which is often constrained by 3D-XPoint write bandwidth. Such mostly-clean caching techniques, which limit the fraction of DRAM cache that can be dirty, hamper the ability for the DRAM cache to act as an effective write buffer for 3D-XPoint. This write limit can cause corresponding degradation in performance, endurance, and power.

Our approach, on the other hand, does not impose any limitation on which lines of the DRAM cache can be kept dirty. Instead, we fundamentally target dirty-bit update cost with architectural techniques. Our DRAM Cache Dirtiness and Preemptive Dirty Marking techniques reduce over 90\% of the bandwidth cost to track dirty information, while needing only 34KB of SRAM storage.


\subsection{DRAM + NVM Hybrid Memories}

There has been a long line of work on hybrid DRAM + NVM systems~\cite{hybridpcm_1,hybridpcm_2,hybridpcm_3}. These works typically try to use DRAM to hide the 4-8x read latency and poor write characteristics of NVM (e.g., low write bandwidth, high power consumption, low write endurance). Our work follows on this line of research.
We develop a scalable (low 34KB SRAM cost) and bandwidth-efficient DRAM cache design, and add a NVM-specific Write-Aware Bypassing that specifically targets hiding NVM's poor write-relative-to-read characteristics.


\section{Conclusion}

This paper investigates bandwidth-efficient DRAM caching for hybrid DRAM + 3D-XPoint memories. 
Effective DRAM caching in front of 3D-XPoint is critical to enabling a memory system that has the apparent high-capacity of 3D-XPoint, and the low-latency and high-write-bandwidth of DRAM. There are two currently major approaches for DRAM cache design:
(1) a Tag-Inside-Cacheline (TIC) organization that optimizes for hits, by storing tag next to each line such that one access gets both tag and data, and (2) a Tag-Outside-Cacheline (TOC) organization that optimizes for misses, by storing tags from multiple data lines together in a tag-line such that one access to a tag-line gets information on several data-lines. Ideally, we would like to have the low hit-latency of TIC designs, and the low miss-bandwidth of TOC designs. To this end, we propose a \textit{TicToc} organization that provisions both TIC and TOC to get the hit and miss benefits of both.

However, we find that naively combining both techniques actually performs worse than TIC individually, because one has to pay the bandwidth cost of maintaining both metadata.
We find the majority of update bandwidth is due to maintaining the TOC dirty information.
We propose \textit{DRAM Cache Dirtiness Bit} that helps prune repeated dirty-bit updates for known dirty lines. We propose \textit{Preemptive Dirty Marking} technique that predicts which lines will be written and proactively marks the dirty bit at install time, to help avoid even the initial dirty-bit update for dirty lines. To support PDM, we develop a novel PC-based \textit{Write-Predictor} to aid in marking only write-likely lines. 
Our evaluations on a 4GB DRAM cache in front of 3D-XPoint show that our TicToc organization enables 10\% speedup over the baseline TIC, nearing the 14\% speedup possible with an idealized DRAM cache design with 64MB of SRAM tags, while needing only 34KB SRAM.

\ignore{
This paper investigates bandwidth-efficient DRAM caching for hybrid DRAM + 3D-XPoint memories. 
3D-XPoint is becoming a viable alternative to DRAM as it enables high-capacity and non-volatile main memory systems. 
However, 3D-XPoint has several characteristics that limit it from outright replacing DRAM: 4-8x slower read, and even worse writes. As such, effective DRAM caching in front of 3D-XPoint is important to enable a high-capacity, low-latency, and high-write-bandwidth memory. There are currently two major approaches for DRAM cache design:
(1) a Tag-Inside-Cacheline (TIC) organization that optimizes for hits, by storing tag next to each line such that one access gets both tag and data, and (2) a Tag-Outside-Cacheline (TOC) organization that optimizes for misses, by storing tags from multiple data lines together in a tag-line such that one access to a tag-line gets information on several data-lines. Ideally, we would like to have the low hit-latency of TIC designs, and the low miss-bandwidth of TOC designs. To this end, we propose a \textit{TicToc} organization that provisions both TIC and TOC to get the hit and miss benefits of both.

However, we find that naively combining both techniques actually performs worse than TIC individually, because one has to pay the bandwidth cost of maintaining both metadata.
The main contribution of this work is developing architectural techniques to reduce bandwidth cost to access and maintain both TIC and TOC metadata. 
We find the majority of update bandwidth is due to maintaining the TOC dirty information.
We propose a \textit{DRAM Cache Dirtiness Bit} technique that carries DRAM cache dirty information to last-level caches, to help prune repeated dirty-bit updates for known dirty lines. 
We propose a \textit{Preemptive Dirty Marking} technique that predicts which lines will be written and proactively marks the dirty bit at install time, to help avoid even the initial dirty-bit update for dirty lines. To support PDM, we develop a novel PC-based \textit{Write-Predictor} to aid in marking only write-likely lines. 
Our evaluations on a 4GB DRAM cache in front of 3D-XPoint show that our TicToc organization enables 10\% speedup over the baseline TIC, nearing the 14\% speedup possible with an idealized DRAM cache design with 64MB of SRAM tags, while needing only 34KB SRAM.
}

\bibliographystyle{ieeetr}

\bibliography{references}

\end{document}